\documentclass[a4paper,10pt,twocolumn]{article}
\usepackage{graphicx}
\usepackage{xcolor}
\usepackage{url}
\usepackage[colorlinks=true,linkcolor=blue,citecolor=blue]{hyperref}
\usepackage[expproduct=cdot]{siunitx} 

\usepackage{amsmath}
\usepackage{bbold} 
\usepackage[charter]{mathdesign} 

\usepackage{ctable}\newcommand{\thickmidrule}{\midrule[\heavyrulewidth]} 
\usepackage{tabularx} 

\usepackage{cite} 

\definecolor{royalblue4}{HTML}{27408B}
\definecolor{red4}{HTML}{8B0000}
\definecolor{green4}{HTML}{008b00} 

\usepackage{dblfloatfix} 
\usepackage{fixltx2e} 

\usepackage[font=footnotesize,labelfont=bf,labelsep=endash,margin=0pt]{caption} 
\usepackage[title,header]{appendix}

\newlength{\myleftmargin} 
\newlength{\mytopmargin} 
\newlength{\myrightmargin} 
\newlength{\mybottommargin} 
\usepackage[runin]{abstract}

\newcommand{\keywords}[1]{\vspace{2mm}\noindent\textbf{Key words:} #1} 
\newcommand{\pagewidetitle}[3] 
{%
    \twocolumn%
        [%
            \vskip-5mm%
            \begin{@twocolumnfalse}%
                #1%
                #2%
                \vspace{5mm}%
            \end{@twocolumnfalse}%
        ]%
        #3%
}

\newlength{\figurewidth}

\newcommand{\ie}{{i.e.}}

\newcommand{\eg}{{e.g.}}
\newcommand{\etal}{{et\,al.}}

\renewcommand{\d}{\mathrm{d}}

\newcommand{\yr}{\ensuremath{\text{yr}}}
\newcommand{\da}{\ensuremath{\text{day}}} 

\newcommand{\um}{\ensuremath{\micro\metre}}
\newcommand{\pM}{\ensuremath{\text{pM}}} 

\usepackage{relsize} 
\newcommand{\bmu}{\textsc{bmu}} 

\newcommand{\ob}{\textsc{ob}}
\newcommand{\oc}{\textsc{oc}}

\newcommand{\obu}{\text{\textsc{ob}$_\text{u}$}} 

\newcommand{\obp}{\text{\textsc{ob}$_\text{p}$}}

\newcommand{\oba}{\text{\textsc{ob}$_\text{a}$}}

\newcommand{\ocu}{\text{\textsc{oc}$_\text{u}$}}
\newcommand{\ocp}{\text{\textsc{oc}$_\text{p}$}}

\newcommand{\oca}{\text{\textsc{oc}$_\text{a}$}}

\newcommand{\tgfb}{\textsc{tgf\textsmaller{$\betaup$}}}
\newcommand{\Tgfb}{\textsc{Tgf\textsmaller{$\betaup$}}}

\newcommand{\wnt}{\text{\textsc{w}\textsmaller{nt}}}

\newcommand{\mcsf}{\textsc{mcsf}}
\newcommand{\rank}{\textsc{rank}}
\newcommand{\rankl}{\textsc{rankl}}
\newcommand{\Rankl}{\textsc{Rankl}}
\newcommand{\opg}{\textsc{opg}}

\def\pgde2{\text{\textsc{pgde}$_2$}}
\newcommand{\pth}{\textsc{pth}}

\newcommand{\bm}{\text{bm}}
\newcommand{\vas}{\text{vas}}
\newcommand{\kform}{\text{$k_\text{form}$}} 
\newcommand{\kres}{\ensuremath{k_\text{res}}}

\newcommand{\sed}{{\ensuremath{\psi_\bm}}}
\newcommand{\sedst}{{\ensuremath{\psi_\bm(t_0)}}}
\newcommand{\fvas}{{\ensuremath{f_\vas}}}
\newcommand{\fvasmax}{{\ensuremath{f_\vas^\ast}}}
\newcommand{\fbm}{{\ensuremath{f_\bm}}}
\newcommand{\sv}{{\ensuremath{S_V}}}

\begin{document}
    
\title{\textbf{The influence of bone surface availability in bone remodelling---A mathematical model including coupled geometrical and biomechanical regulations of bone cells}}
\author{Peter Pivonka$^\text{a,1}$, Pascal R Buenzli$^\text{a}$, Stefan Scheiner$^\text{a,b}$, Christian Hellmich$^\text{b}$, Colin R Dunstan$^\text{c}$}

\date{\small%
    \vspace{-2mm}%
    $^\text{a}$Engineering Computational Biology Group, The University of Western Australia, Perth WA 6009, Australia%
    \\$^\text{b}$Institute for Mechanics of Materials and Structures, Vienna University of Technology, A-1040 Vienna, Austria%
    \\$^\text{c}$Mechanical \& Mechatronic Engineering, University of Sydney, Sydney NSW 2006, Australia%
    \\\vskip 1mm \normalsize \today\vspace*{-5mm}}

\pagewidetitle{
\maketitle}{
\begin{abstract}
    Bone is a biomaterial undergoing continuous renewal. The renewal process is known as bone remodelling and is operated by bone-resorbing cells (osteoclasts) and bone-forming cells (osteoblasts). An important function of bone remodelling is the repair of microcracks accumulating in the bone matrix due to mechanical loading. Cell-cell communication between cells of the osteoclastic lineage and cells of the osteoblastic lineage is thought to couple resorption and formation so as to preserve bone integrity and achieve homeostatic bone renewal. Both biochemical and biomechanical regulatory mechanisms have been identified in this coupling. Many bone pathologies are associated with an alteration of bone cell interactions and a consequent disruption of bone homeostasis. In osteoporosis, for example, this disruption leads to long-term bone loss and fragility, and can ultimately result in fractures.

Here we focus on an additional and poorly understood potential regulatory mechanism of bone cells, that involves the morphology of the microstructure of bone. Bone cells can only remove and replace bone at a bone surface. However, the microscopic availability of bone surface depends in turn on the ever-changing bone microstructure. The importance of this geometrical dependence is unknown and difficult to quantify experimentally. Therefore, we develop a sophisticated mathematical model of bone cell interactions that takes into account biochemical, biomechanical and geometrical regulations. We then investigate numerically the influence of bone surface availability in bone remodelling within a representative bone tissue sample. Biochemical regulations included in the model involve signalling molecules of the receptor--activator nuclear factor $\kappa$B pathway (\rank--\rankl--\opg), macrophage colony-stimulating factor (\mcsf), transforming growth factor~$\betaup$ (\tgfb) and parathyroid hormone (\pth). For the biomechanical regulation of bone cells, a multiscale homogenisation scheme is used to determine the microscopic strains generated at the level of the extravascular matrix hosting the osteocytes by macroscopic loading. The interdependence between the bone cells' activity, which modifies the bone microstructure, and changes in the microscopic bone surface availability, which in turn influences bone cell development and activity, is implemented using a remarkable experimental relationship between bone specific surface and bone porosity. Our model suggests that geometrical regulation of the activation of new remodelling events could have a significant effect on bone porosity and bone stiffness. On the other hand, geometrical regulation of late stages of osteoblast and osteoclast differentiation seems less significant. We conclude that the development of osteoporosis is probably accelerated by this geometrical regulation in cortical bone, but probably slowed down in trabecular bone.

\keywords{mechanical feedback, geometrical feedback, specific surface, bone remodelling, bone stiffness, osteoporosis}
\end{abstract}
}{
\protect\footnotetext[1]{Corresponding\mbox{ }author.\mbox{ }Email\mbox{ }addresses: \texttt{peter.pivonka@uwa.edu.au}\mbox{ }(Peter\mbox{ }Pivonka), \texttt{pascal.buenzli@uwa.edu.au}\mbox{ }(Pascal\mbox{ }R\mbox{ }Buenzli), \texttt{stefan.scheiner@tuwien.ac.at}\mbox{ }(Stefan\mbox{ }Scheiner), \texttt{christian.hellmich@tuwien.ac.at}\mbox{ }(Christian\mbox{ }Hellmich), \texttt{c.dunstan@usyd.edu.au}\mbox{ }(Colin\mbox{ }R\mbox{ }Dunstan).}
}

\section{Introduction}
Bone is a biomaterial that has a variety of physiological functions. In addition to load bearing and support for locomotion, bone protects internal organs and participates in calcium and phosphorous homeostasis. From an engineering perspective the structural function of bone is most importantly characterised by its stiffness and strength. Daily activities (such as walking and running) subject bone to periodical loads which, over extended periods of time (weeks, months and years), can lead to fatigue damage and the formation of microcracks. If these microcracks are not removed in due time, their evolution may result in a macroscopic structural failure, \ie, a fragility fracture. To prevent the occurrence of fatigue fractures, nature has equipped bone tissues with a cellular mechanism of self-repair~\cite{taylor-hazenberg-lee}, referred to by biologists as `bone remodelling'~\cite{parfitt2-in-recker,martin-burr-sharkey}. Bone remodelling is a coordinated process of bone resorption by cells called osteoclasts, and bone formation by cells called osteoblasts. Osteoclasts and osteoblasts usually operate together in self-contained groups processing the renewal of a localised portion of the bone tissue. These groups are called bone multicellular units (\bmu s) and constitute a single `remodelling event'. There are about \num{1.7e6} such \bmu s in a normal adult skeleton~\cite{parfitt-1994,martin-burr-sharkey,parfitt2-in-recker}. Cell population and cell activity in a \bmu\ are tightly controlled to establish local bone homeostasis (\ie, balanced bone resorption and bone formation). In bone pathologies, this cellular control is perturbed and homeostatic bone renewal is disrupted. In osteoporosis, bone is progressively lost, which results in reduced bone stiffness and strength.

Over the last decades, bone biologists have identified a large number of biochemical regulatory factors influencing bone remodelling. The formation of osteoclasts has been shown to rely crucially on macrophage colony-stimulating factor (\mcsf) and on the receptor-activator nuclear factor $\kappa$B (\rank) cell signalling pathway, which involves the receptor \rank, the ligand \rankl\ and osteoprotegerin (\opg)~\cite{martin-rankl,roodman}. \Rankl\ activates the \rank\ receptor on precursor osteoclasts, which triggers their development and sustains their activity. The soluble molecule \opg\ is a decoy receptor of \rankl\ which can prevent \rankl\ from binding to \rank. Another important molecule mediating the communication between osteoblasts and osteoclasts is transforming growth factor $\betaup$ (\tgfb). \Tgfb\ is stored in high concentrations in the bone matrix. It is released into the bone microenvironment, where it exerts its action on several bone cells, during bone matrix resorption by active osteoclasts~\cite{roodman}. The existence of a mechanical regulation of bone remodelling has long been suspected. It is now well established that mechanical feedback is a key regulatory mechanism to maintain bone mass \cite{frost-mechanostat1,frost-mechanostat2,frost-mechanostat3,smit-burger,burger-klein-nulend-smit,lee-staines-taylor}. The commonly accepted view is that osteocytes act as mechanosensors that transduce local mechanical signals into biochemical responses. These biochemical responses are thought to regulate the initiation of bone remodelling processes and to modulate the coupling between bone resorption and formation (see \eg~\cite{taylor-hazenberg-lee} and Refs\ cited therein).

The existence of biochemical and biomechanical regulations of bone cells is well-established and has been extensively studied. However, the notion that the morphology of the microstructure of bone may induce an additional regulation of bone cells of purely geometrical nature is not often mentioned in the recent literature. This may be due to the experimental difficulty of assessing the importance of a geometrical regulation. Biochemical and biomechanical regulations can experimentally be partially or fully repressed by selective gene knock-outs or monoclonal antibodies targeting key components in the bone cell signalling pathways. By contrast, one cannot simply ``switch off'' a geometrical regulation of bone remodelling when this self-repair process modifies the microstructure (and so the geometry) of the material.

\begin{figure}[t]
    \centering    \makebox[0pt]{\hspace{90mm}\raisebox{0.5mm}[0pt][0pt]{\includegraphics[width=\figurewidth]{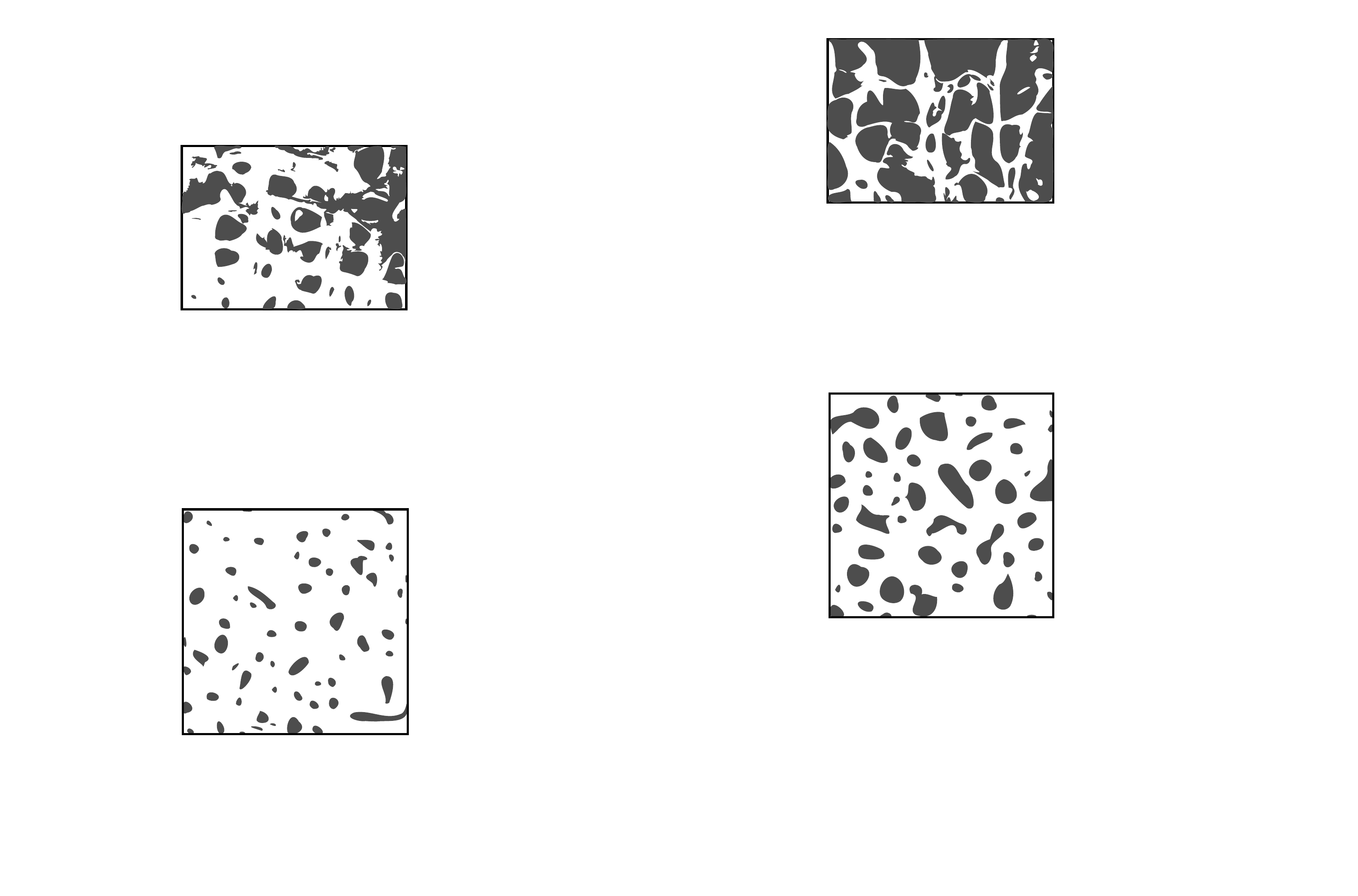}}}%
\includegraphics[width=\figurewidth]{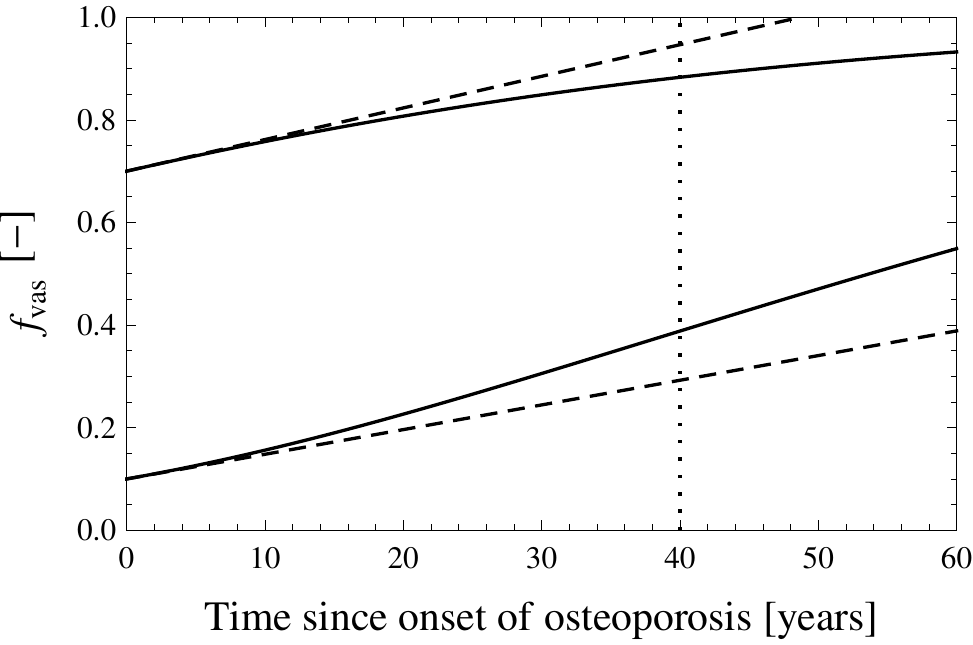}
        \caption{Schematic representation of a possible effect of geometrical feedback on the evolution of vascular porosity (\fvas) in osteoporosis both in cortical bone (lower curves) and trabecular bone (upper curves) according to Martin~\cite{martin-1984}. Dashed curves show the linear increase in porosity that is obtained without geometrical feedback, while the solid curves incorporate geometrical regulation (a constant pathological skeletal imbalance of -2~\um/\yr\ is assumed). Typical bone microstructures for cortical bone (bottom) and trabecular bone (top) in normal subjects (left) and osteoporotic subjects (right) are also represented.}
    \label{fig:martin}
\end{figure}

Bone tissue is diverse and exhibits a broad variety of microstructures. However, two distinctive types of bone tissue are usually identified: cortical bone and trabecular bone~\cite{martin-burr-sharkey} (see images in Figure~\ref{fig:martin}). Cortical bone has typical porosities of $0.05$--$0.15$ while trabecular bone has typical porosities of $0.65$--$0.85$~\cite{martin-burr-sharkey,parfitt2-in-recker}. Mathematical models for the estimation of mechanical properties of bone tissue have shown that bone stiffness is predominantly determined by the porosity $\fvas$,\footnote{\label{footnote:canaliculi-porosity}The total porosity of bone is made of a vascular porosity, which contains marrow components, blood vessels, bone cells and their precurors, and the lacunae-canaliculi porosity, which contains osteocytes and their processes. The lacunae-canaliculi porosity is only a small fraction of the total porosity (see \eg~\cite[Table 1]{martin-1984}) and no remodelling occurs at these surfaces. Therefore, the lacunae-canaliculi porosity will not be consider in this work and we will refer to the vascular porosity simply as the bone porosity. Similarly, in the present context, we are not interested in the intercrystalline and intermolecular porosities, which we simply regard as part of the `solid bone matrix'.} the interaction of the different material phases and pore shape, while other microstructural characteristics such as the exact pore distribution play a secondary role~\cite{hellmich-etal,fritsch-hellmich,grimal-etal}. For biochemical processes, pore morphology can be expected to play a significant role. Indeed, pore morphology determines the so-called specific surface $\sv$ (\ie, the amount of bone surface available in a representative volume element), which is an essential geometrical factor for the bone cells. Bone cells require a bone surface to fulfill their functions, whether to initiate a bone remodelling process or to operate resorption and formation. Osteoclasts require attachment to a particular area of the bone surface before resorbing. Osteoblasts are observed to only secrete osteoid (a collagen-rich substance which later mineralises and becomes new bone matrix) on existing bone surfaces. Finally, mechanical signals sensed by osteocytes embedded in the bone matrix are passed on to bone cells in the vascular cavity through the bone surface. Effects similar to chemical exchange reactions between pore walls and solutes in fluid-saturated porous materials can be expected to occur in this context. 

The issue of quantifiying the role of bone surface availability in bone remodelling was raised by bone biologists already some time ago~\cite{martin-1972,martin-1984,parfitt2-in-recker}. In Ref.~\cite{martin-1984}, Martin provides a first attempt to investigate theoretically the effect of a geometrical regulation of bone remodelling in osteoporosis (see Figure~\ref{fig:martin}). Osteoporosis is associated with increased porosity in both cortical and trabecular bone~\cite{nishiyama-etal}. In Martin's own words: \emph{``In [cortical] bone, increased porosity provides more surface area on which cells can work, thereby increasing the capacity for further porosity changes. In [trabecular] bone, increased porosity decreases the amount of surface available to the cells, thereby decreasing the capacity for further remodelling.''}

While the proposed mechanism of geometrical feedback on bone remodelling seems plausible, it is difficult to test its validity experimentally and to determine its importance quantitatively. Some researchers have employed the concept of geometrical feedback for simulations of bone remodelling~\cite{hazelwood-etal}. However, to our knowledge, there is no systematic study in the literature of the effects of a possible geometrical regulation at several stages of the remodelling sequence. Also, the interplay between geometrical feedback and mechanical feedback in bone remodelling has not been investigated. A mechanical feedback has the potential to stabilise bone loss or gain \cite{frost-mechanostat1,scheiner-etal-mechanostat} and may either compete with or enhance the effect of the geometrical feedback seen in Figure~\ref{fig:martin} depending on the type of bone.

The aim of this paper is to address the above questions using a state-of-the-art computational model of bone remodelling. The following questions related to geometrical feedback are investigated: (i) At which stage of the bone remodelling sequence (\ie, activation, resorption, formation) does geometrical feedback have the strongest effect?; (ii) How do geometrical and mechanical feedbacks interact?; (iii) What is the impact of geometrical feedback in osteoporosis in terms of bone porosity and bone stiffness?

To address these questions we extend a previously developed mathematical model of bone remodelling~\cite{pivonka-etal-1,pivonka-etal-2,scheiner-etal-mechanostat}. This multiscale model takes into account both biochemical and biomechanical regulations of bone remodelling. Biochemical regulatory factors include the \rank--\rankl--\opg\ pathways together with the action of \tgfb\ on bone cells. Biomechanical regulation of bone formation and bone resorption is mediated by the microscopic strain energy density (\textsc{sed}) of the bone matrix. This strain energy density is calculated from a micromechanical homogenisation scheme. The introduction of geometrical feedback due to microscopic bone surface availability is elucidated through a phenomenological relationship between the specific surface and the vascular porosity obtained from various types of bone~\cite{martin-1984}.

\section{Mathematical model of bone remodelling}\label{sec:model}
Few mathematical models of bone remodelling include explicitly biochemical interactions of bone cells that couple bone resorption and bone formation. Lemaire \etal~\cite{lemaire-etal} have proposed a bone cell population model that incorporates some of the most important known bone biology. This cell population model was further developed by Pivonka \etal\ to investigate the effect of \rankl\ expression on osteoblasts of varying maturity~\cite{pivonka-etal-1}. These models have been shown to correctly capture important physiological behaviours of bone remodelling both in bone homeostasis and in bone pathologies~\cite{pivonka-etal-2,buenzli-pivonka-smith,wang-etal-bone-mm,scheiner-etal-denosumab}. Recently, we have proposed an extension of the model of Ref.~\cite{pivonka-etal-1} to include a biomechanical regulatory mechanism to trigger bone cell responses, leading to a fully coupled model of biochemical and biomechanical regulations~\cite{scheiner-etal-mechanostat}. This model uses a novel multiscale approach based on micromechanical homogenisation of bone stiffness. It allows to consistently calculate bone matrix strains at the osteocyte level. Osteocytes convert this micromechanical signal into biochemical signals to bone cells in the vascular space, effectively providing a biomechanical feedback regulation of bone remodelling.

In the following, we present an extension of the model of Ref.~\cite{scheiner-etal-mechanostat} that incorporates the influence of microscopic bone surface availability at various stages of the bone remodelling sequence.

\subsection{Fundamental biochemical and biomechanical regulation of bone remodelling}\label{sec:biochem-biomech-regulation-biology}
\begin{figure*}
    \centering\includegraphics[width=0.65\textwidth]{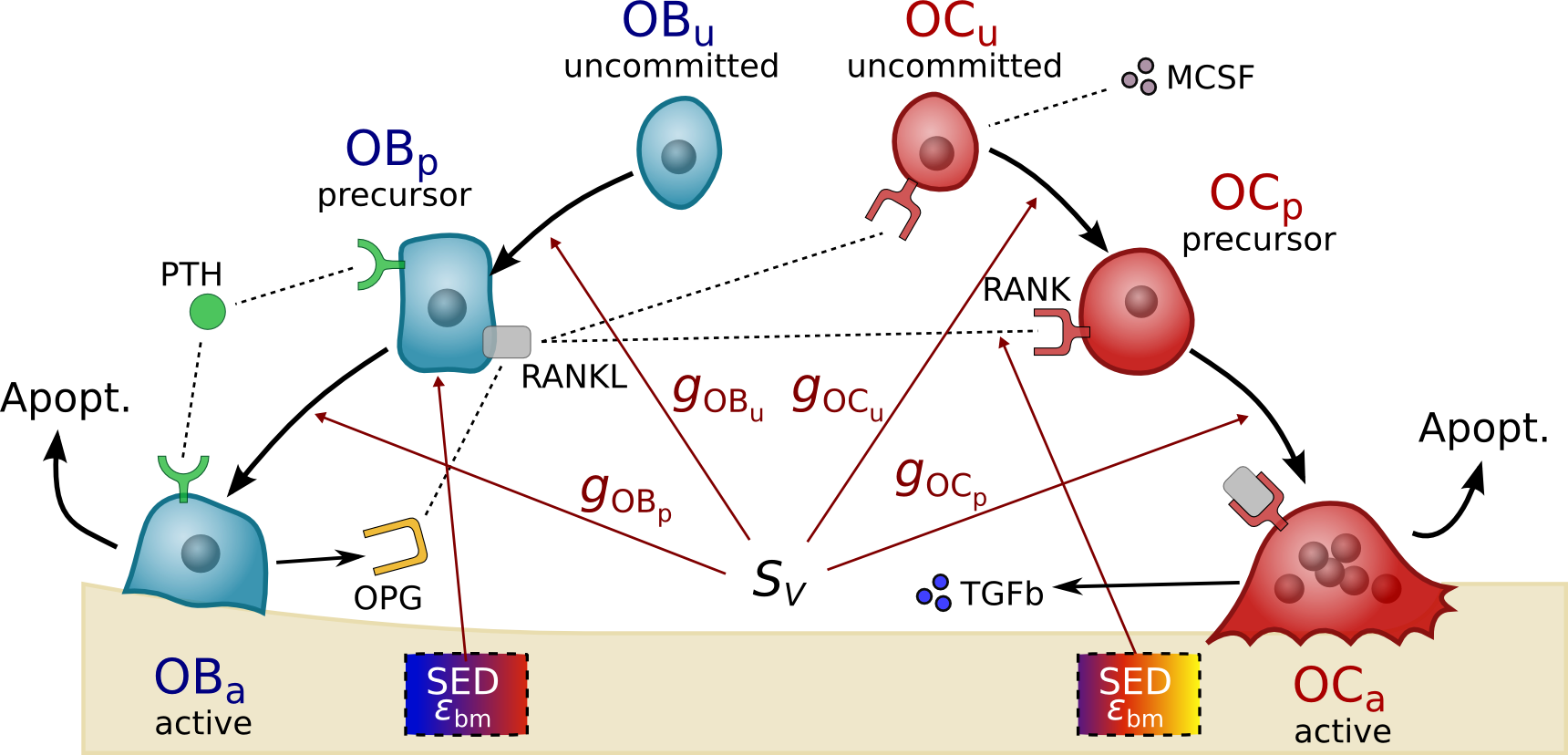} 
    \caption{Cell population model of bone remodelling, including several developmental stages of osteblasts (\obu, \obp, \oba) and osteoclasts (\ocu, \ocp, \oca) and their biochemical regulation (\pth, \rank--\rankl--\opg, \tgfb, \mcsf), biomechanical regulation (\textsc{sed}, $\varepsilon_\bm$) and geometrical regulation (\sv) (see text for further explanations).}
    \label{fig:model}
\end{figure*}

The biochemical regulatory mechanisms considered in the model have been described in detail in Refs.~\cite{pivonka-etal-1,pivonka-etal-2}. Three stages of osteoblast development are considered: uncommitted osteoblast precursors (\obu s), pre-osteoblasts (\obp s) and active osteoblasts (\oba s). Similarly, three stages of osteoclast development are considered: uncommitted osteoclast precursors (\ocu s), pre-osteoclasts (\ocp s) and active osteoclasts (\oca s). Figure~\ref{fig:model} schematically depicts these bone cell types along with their biochemical, biomechanical and geometrical regulations. These regulations are summarised sequentially in the following.

The generation of osteoblasts is assumed to be regulated by transforming growth factor $\betaup$ (\tgfb) (a factor released from the bone matrix during osteoclastic bone resorption) while the generation of osteoclasts is assumed to be regulated by \rankl\ and \opg\ (molecules in the receptor-activator nuclear factor $\kappa$B (\rank) system, that are expressed by osteoblasts). \Tgfb\ promotes the differentiation of uncommitted osteoblast progenitors (\obu) into pre-osteoblasts (\obp), but it inhibits the differentiation of pre-osteoblasts into active osteoblasts (\oba). Furthermore, \tgfb\ promotes osteoclast apoptosis (programmed cell death). \Rankl\ is a protein expressed on the surface of pre-osteoblasts. The binding of \rankl\ to the receptor \rank\ found on pre-osteoclasts promotes the differentiation of pre-osteoclasts into active osteoclasts. However, this binding may be prevented by the presence of \opg, a decoy receptor of \rankl\ produced in soluble form by active osteoblasts. Furthermore, the circulating parathyroid hormone \pth\ induces \rankl\ expression and downregulates \opg\ expression by osteoblasts to produce a systemic increase in available \rankl, resulting in an increase in the formation and activity of osteoclasts. The differentiation of uncommitted osteoclast progenitors (\ocu s) into pre-osteoclasts (\ocp s) requires signalling by both \rankl\ and macrophage colony-stimulating factor (\mcsf). All these signalling actions are accounted for using mass action kinetics for the chemical bindings between ligands and their receptors~\cite{lauffenburger-linderman}. The fraction of occupied receptors on a cell (\ie, bound to a ligand) is assumed to determine the strength of the signal received by the cell, and so is assumed to determine the strength of the cell's response.

The biomechanical regulatory mechanisms considered in the model are described in detail in Ref.~\cite{scheiner-etal-mechanostat}. Mechanical disuse is known to increase bone resorption by increasing the \rankl/\opg\ ratio, which increases the differentiation of pre-osteoclasts into active osteoclasts. In the model, this is implemented using a mechanically controlled \rankl-production term, see Eq.~\eqref{Prankl}. Mechanical overuse is believed to increase bone formation by stimulating \wnt\ signalling to pre-osteoblasts, which increases their proliferation and ultimately leads to an increased population of active osteoblasts~\cite{scheiner-etal-mechanostat,bonewald-johnson}. In the model, this is implemented using a mechanically controlled proliferation of pre-osteoblasts, see Eq.~\eqref{Pieps}.

\subsection{Geometrical and morphological characteristics of bone: porosity and specific surface}
To represent the microstructure of bone at the tissue level, the most important geometrical and morphological parameters are the vascular porosity (\fvas) and the specific surface (\sv). In cortical bone, vascular porosity corresponds to the so-called `Haversian canal' system. In trabecular bone, vascular porosity corresponds to the marrow space around the trabecular struts~\cite{martin-burr-sharkey}. The vascular compartment contains all the bone cells considered in our model. Vascular porosity is defined as the volume fraction of vascular pores, \ie, the volume of vascular pores ($V_\vas$) per tissue volume ($V_T$):\footnote{\label{footnote:rve}The tissue volume $V_T$ is assumed to be of the order of 1--3~\milli\metre$^3$. This volume corresponds to a representative volume element (RVE) large enough to comprise several remodelling events (\bmu s)~\cite{parfitt2-in-recker}, yet small enough to represent spatial heterogeneity of bone tissue (in particular, small enough to distinguish cortical bone and trabecular bone)~\cite{grimal-etal}.} 
\begin{align}
    \fvas = V_\vas/V_T.\label{fvas-def}
\end{align}
The bone matrix volume fraction is defined in the same way as the volume of bone matrix ($V_\bm$) per tissue volume, \ie:
\begin{align}
    \fbm = V_\bm/V_T.\label{fbm}
\end{align}
We recall that all porosities at observation scales below the vascular porosity, such as the lacunar porosity, and the canaliculi connecting the lacunae, are not involved in the remodelling process, i.e. the intricate biochemical processes take place within the vascular pore space (see also footnote~\ref{footnote:canaliculi-porosity}). From Eqs~\eqref{fvas-def}--\eqref{fbm}, it follows that $\fvas + \fbm = 1$. Typically, cortical bone exhibits a range of porosities $\fvas\approx 0.05$--$0.15$ while trabecular bone exhibits a range of porosities $\fvas\approx 0.65$--$0.85$.

The specific surface (or surface density) of a porous material is defined as the interstitial surface area of the pores ($S_p$) per tissue volume, \ie:
\begin{align}
    \sv = S_p/V_T
\end{align}
with dimensions $[\mm^2/\mm^3]$. Generally speaking, the specific surface is an important quantity for a variety of phenomena in porous media. For example, it determines the adsorption capacity of industrial adsorbents and plays an important role in determining the effectiveness of catalysts and ion exchange columns and filters. It is also related to the fluid conductivity or permeability of porous media (see \eg\ Ref.~\cite{dullien}). Experimentally, $\sv$ is commonly estimated by adsorption methods, quantitative stereology,\footnote{Stereology is the study of three-dimensional properties of objects observed in two-dimensional sections.} fluid flow and micro-computed tomography. As mentioned above, in bone the specific surface is important as it determines the available working area for osteoblasts and osteoclats. The specific surface can also be expected to have an influence on the transmission of specific signalling by osteocytes in the bone matrix to osteoblasts and osteoclasts developing in the vascular pores.

\begin{figure}[t]
    \centering
    \includegraphics[width=\figurewidth]{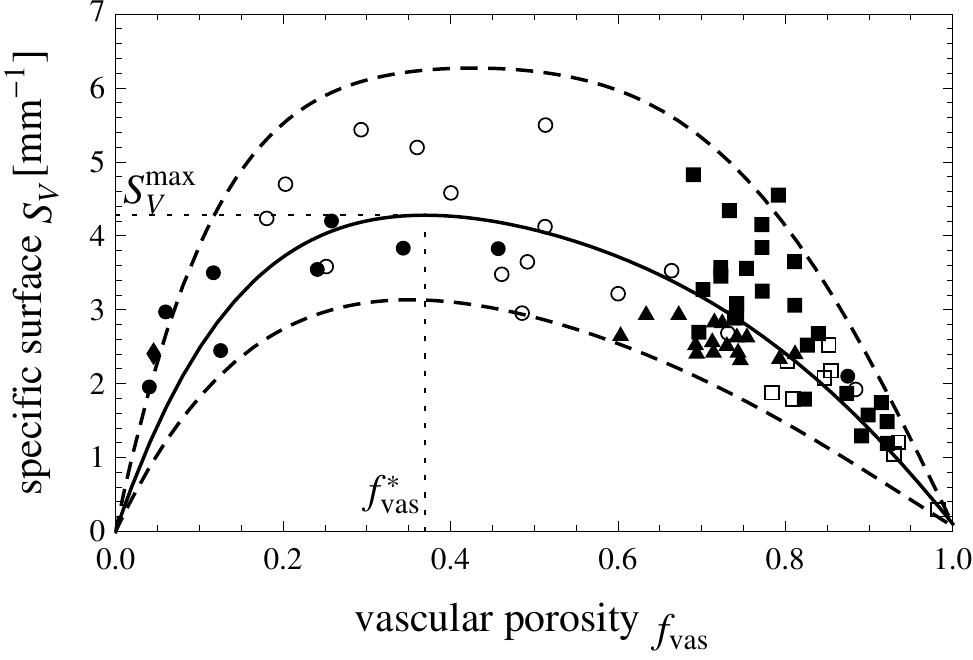}
    \caption{Relation between bone specific surface (\sv) and vascular porosity (\fvas) (modified from Martin~\cite[Fig.~19]{martin-1984}). Data points were measured on histological sections and microradiographs of different types of bone. Solid symbols: healthy bone; open symbols: diseased bone (osteoporosis, osteogenesis imperfecta, and osteomalacia); circles: human femoral bone; squares: human iliac crest; diamonds: human rib; triangles: human vertebra. The maximum of the curve $\sv(\fvas)$ (thick solid line, Eq.~\eqref{sv-fvas}) at $\fvas = \fvasmax \approx 0.37$ is indicated by dotted lines. The dashed curves are obtained by varying the polynomial coefficients in Eq.~\eqref{sv-fvas} to enclose most of the range of measurements.}\label{fig:sv-fvas}
\end{figure}

The microstructure of a material determines both the porosity and the specific surface. Depending on the microstructure, different materials exhibit different values for these quantities. Bone tissue spans a wide range of porosities, each of which is characteristic of a particular micro-architecture, and so of a particular value of the specific surface.  Based on a large number of experimental data, Martin has provided a remarkable phenomenological relationship between bone specific surface (\sv) and vascular porosity (\fvas)~\cite[Eq.~(68)]{martin-1984}:
\begin{align}
    \sv(\fvas) = a\, \fvas + b\, \fvas^2 + c\,\fvas^3 + d\,\fvas^4 + e\,\fvas^5,\label{sv-fvas}
\end{align}
where the polynomial coefficients are estimated as $a=32.3, b=-93.9, c=134, d=-101$, and $e=28.8$ (in [$\mm^{-1}$]). In Figure~\ref{fig:sv-fvas}, the relation~\eqref{sv-fvas} is plotted together with experimental data obtained from various types of human bone (femur, iliac crest, vertebra, rib) both in health and disease. The data and the curve $\sv(\fvas)$ show two important characteristics: (i)  all specimens approximately follow the same $\sv(\fvas)$ curve independently of bone type and with no significant difference between diseased and healthy bone. This remarkable universality establishes the curve $\sv(\fvas)$ as an intrinsic property of bone; (ii) the specific surface exhibits a maximum at a porosity $\fvasmax$ of about  0.37, intermediate between cortical and trabecular bone (bone at such porosity is denoted as `porous-compact' bone).

In this work, we use Eq.~\eqref{sv-fvas} to include geometrical regulation in the model as follows. The evolution of the bone cell populations predicts the evolution of the vascular porosity due to resorption by osteoclasts and formation by osteoblasts. Eq.~\eqref{sv-fvas} then enables us to estimate changes in the specific surface associated to the changes in porosity. This change in microscopic bone surface availability is in turn assumed to influence the evolution of the cell populations.

\subsection{Bone cell governing equations}
In the model, osteoblasts and osteoclasts of different developmental stages are considered. The populations of osteoblasts and osteoclasts are therefore heterogeneous. The composition of these populations is determined by following individually the populations of each of the developmental stages of osteoblasts and osteoclasts mentioned above. The evolution of these cell (sub)populations is transcribed mathematically as so-called `rate equations'~\cite{pivonka-etal-1,scheiner-etal-mechanostat}. The populations of uncommitted progenitor osteoblasts (\obu s) and uncommitted progenitor osteoclasts (\ocu s) are assumed constant and so are not state variables. These uncommitted cells represent a pool of progenitor (or stem) cells that is assumed to be maintained by self-renewal unlimitedly. The possibility for geometrical regulation is included at each stage of osteoblast and osteoclast development through functions of the specific surface, namely $g_\obu(\sv)$, $g_\obp(\sv)$, $g_\ocu(\sv)$ and $g_\ocp(\sv)$ (see Figure~\ref{fig:model}). In the following, we denote the bone cell densities within a tissue sample (number of cells per unit volume) by their symbol \obu, \obp, \oba, \ocu, \ocp, \oca. The concentrations of the biochemical signalling molecules within a tissue sample (number of molecules per unit volume) is also denoted by their symbol \tgfb, \rankl, etc.\footnote{To align with common practice, we shall use the terminology `density' for cells and `concentration' for signalling molecules, even if the units are chosen the same.} Based on the above descriptions of the biochemical, biomechanical, and geometrical regulatory mechanisms, the governing equations of the bone cell densities in the model are expressed as:
\begin{align}
    \frac{\d}{\d t} \obp &= \Big\{g_\obu(\sv)\, D_\obu\, \pi^\tgfb_{\text{act},\obu}\Big\}\, \obu + \Big\{P_\obp\, \Pi^{\,\sed}_{\text{act},\obp}\Big\}\, \obp\notag
    \\&\phantom{=\ } - \Big\{g_\obp(\sv)\, D_\obp\, \pi^\tgfb_{\text{rep},\obp}\Big\}\, \obp\label{obp}
    \\\frac{\d}{\d t} \oba &= \Big\{g_\obp(\sv)\, D_\obp\, \pi^\tgfb_{\text{rep},\obp}\Big\}\, \obp - A_{\oba}\, \oba \label{oba}
    \\\frac{\d}{\d t} \ocp &= \Big\{g_\ocu(\sv)\, D_\ocu\, \pi^\mcsf_{\text{act}, \ocu}\, \pi^\rankl_{\text{act},\ocu}\Big\}\, \ocu \notag
    \\&\phantom{=\ }- \Big\{g_\ocp(\sv)\, D_\ocp\, \pi^\rankl_{\text{act},\ocp}\Big\}\, \ocp \label{ocp}
    \\\frac{\d}{\d t} \oca &= \Big\{g_\ocp(\sv)\, D_\ocp\, \pi^\rankl_{\text{act},\ocp}\Big\}\, \ocp - \Big\{A_\oca\, \pi^\tgfb_{\text{act}, \oca}\Big\}\, \oca. \label{oca}
\end{align}
Below, we discuss the various quantities occurring in these equations (see Refs~\cite{pivonka-etal-1,scheiner-etal-mechanostat} for more details). The parameters $D_\obu$, $D_\obp$, $D_\ocu$, $D_\ocp$ denote differentiation rate parameters for uncommitted osteoblast progenitors, pre-osteoblasts, uncommitted osteoclast progenitors and pre-osteoclasts, respectively; $P_\obp$ is a proliferation rate parameter for pre-osteoblasts; $A_\oba$ and $A_\oca$ are apoptosis rate parameters for active osteoblasts and active osteoclasts. Biochemical regulation of cell development is achieved through so-called ``activator'' and ``repressor'' functions of the biochemical signalling molecules $\tgfb$, $\mcsf$ and $\rankl$ as follows. The functions $\pi^\tgfb_{\text{act},\oba}$, $\pi^\tgfb_{\text{rep}, \obp}$, and $\pi^\tgfb_{\text{act},\oca}$ are activator and repressor functions regulating osteoblast differentiation and osteoclast apoptosis based on the concentration of \tgfb. The function $\pi^\mcsf_{\text{act},\ocu}$ is an activator function regulating differentiation of \ocu s into \ocp s based on the concentration of macrophage colony stimulation factor (\mcsf). The functions  $\pi^\rankl_{\text{act},\ocu}$ and $\pi^\rankl_{\text{act},\ocp}$ are activator functions regulating osteoclast differentiation based on the concentration of \rankl. For simplicity, we assume that $\pi^\rankl_{\text{act},\ocu}=\pi^\rankl_{\text{act},\ocp}$. The equations governing the evolution of the biochemical signalling molecules \tgfb, \mcsf, \rankl, \rank, \opg\ and \pth\ and the form of the activator and repressor functions are presented in Appendix~\ref{appx:model-description}, see Eqs~\eqref{tgfb}--\eqref{pirankl}. Tables~\ref{table:dynamic-variables}--\ref{table:biomech} in Appendix~\ref{appx:model-description} list the dynamic quantities, the biochemical parameters and the biomechanical quantities of the model.

\subsubsection{Mechanical feedback regulation}
Biomechanical studies suggest that the strain energy density is an important quantity that determines bone adaptation to various mechanical loads. This quantity is commonly chosen in the literature due its scalar nature as a measure of mechanical stimulus sensed by the bone cells to drive bone adaptation~\cite{martin-burr-sharkey,oers-etal}.

\begin{figure}
    \centering\includegraphics[width=\figurewidth]{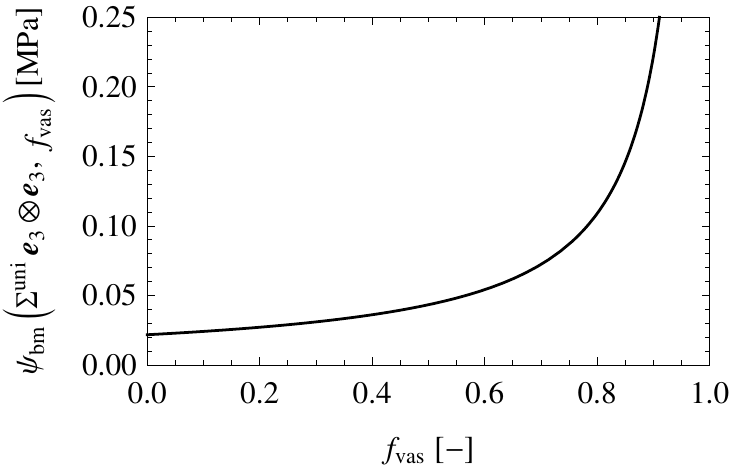}%
    \caption{Microscopic strain energy density \sed\ derived from the micromechanical homogenisation procedure as a function of vascular porosity \fvas\ for the case of a uniaxial compressive loading $\boldsymbol\Sigma=\Sigma^\text{uni} \textbf{e}_3 \otimes \textbf{e}_3$ with $\Sigma^\text{uni} = -30~\text{MPa}$.}
    \label{fig:sed-vs-fvas}
\end{figure}

The model of biomechanical regulation developed in Ref.~\cite{scheiner-etal-mechanostat} similarly uses the strain energy density as the signal conveying mechanical information to the bone cells. However, as discussed in Ref.~\cite{scheiner-etal-mechanostat}, it is paramount to estimate consistently the bone matrix strain energy density at the micro-scale where osteocytes sense this mechanical signal and transduce it into a biochemical response. In Ref.~\cite{scheiner-etal-mechanostat}, we used a homogenisation procedure based on Eshelby's classical matrix inclusion problem and the Mori-Tanaka scheme to estimate the microscopic strains $\varepsilon_\bm$ generated at the level of the extravascular matrix hosting the osteocytes, by the macroscopic loading of the tissue. This homogenisation procedure leads to an intricate dependence of the microscopic strain energy density of the bone matrix, $\psi_\bm$, upon the macroscopic stress tensor $\boldsymbol\Sigma$ and the vascular porosity $\fvas$:
\begin{align}
    \psi_\bm = \psi_\bm(\boldsymbol\Sigma, \fvas).\label{sed}
\end{align}
In this paper, we will exemplify the effect of biomechanical regulation for the situation of a constant uniaxial compressive loading only (\ie, $\boldsymbol\Sigma=\Sigma^\text{uni} \textbf{e}_3\otimes\textbf{e}_3$ with $\Sigma^\text{uni} = -30~\text{MPa}$). For this situation, the dependence of \sed\ upon \fvas\ is plotted in Figure~\ref{fig:sed-vs-fvas}. The reader is referred to Refs~\cite{scheiner-etal-mechanostat,hellmich-etal,fritsch-hellmich,hellmich-kober-erdmann} for the general mathematical specification of \sed\ and a detailed presentation of the homogenisation procedure.
As mentioned in Section~\ref{sec:biochem-biomech-regulation-biology}, the biomechanical regulation of bone remodelling is believed to operate through different pathways for resorption and formation. Following Ref.~\cite{scheiner-etal-mechanostat}, the biomechanical regulation of bone resorption is realised in the model via modulation of the \rank--\rankl--\opg\ signalling pathway by the microscopic strain energy density \sed. The production rate of \rankl\ on pre-osteoblasts, $P_\rankl^{\,\sed}$, is assumed to be enhanced during mechanical disuse:
\begin{align}
    P_\rankl^{\,\sed} =
        \begin{cases}
            \kappa\ \Bigg(1 - \frac{\displaystyle \sed}{\displaystyle \sedst} \Bigg), &\sed< \sedst
            \\0,&\sed\geq \sedst
        \end{cases}
    \label{Prankl}
\end{align}
where $\kappa$ is a parameter quantifying the strength of the biomechanical transduction and $\sedst$ is the steady-state value of the strain energy density. The steady state is assumed to be an initial homeostatic state of bone remodelling, with no bone gain or loss. The increase in \rankl\ production rate during mechanical disuse increases both $\pi^\rankl_{\text{act},\ocu}$ and $\pi^\rankl_{\text{act},\ocp}$ in Eqs~\eqref{ocp} and \eqref{oca}, and consequently leads to increased osteoclast generation (see Appendix~\ref{appx:model-description}, Eqs.~\eqref{rankl}, \eqref{pirankl}).

Following Ref.~\cite{scheiner-etal-mechanostat}, the biomechanical regulation of bone formation is realised in the model via modulation of pre-osteoblast proliferation by the microscopic strain energy density \sed. In Eq.~\eqref{obp}, pre-osteoblasts are generated both by differentiation from \obu s and by self-expansion through proliferation. The modulation of proliferation by the strain energy density is expressed by an `activator' function $\Pi^{\,\sed}_{\text{act}, \obp}$, defined as:
\begin{align}
    \Pi^{\,\sed}_{\text{act}, \obp} = \begin{cases}
        \displaystyle\frac{1}{2}\rule[-2ex]{0pt}{0ex}
        & \sed \leq \sedst,
        \\\displaystyle \frac{1}{2} + \frac{\lambda}{2} 
    \Bigg(\frac{\displaystyle \sed}{\displaystyle \sedst}-1\Bigg), & \sedst < \sed < \sed^\ast,
    \\1 & \sed^\ast \leq \sed,
    \end{cases}
    \label{Pieps}
\end{align}
where $\lambda$ is a constant parameter quantifying the strength of the biomechanical transduction and, $\sed^\ast = (1+\lambda^{-1})\sedst$ is the minimum value of the strain energy density for which \mbox{$\Pi^{\,\sed}_{\text{act}, \obp}=1$}.

\subsubsection{Geometrical feedback regulation}\label{sec:geom-regulation}
The four regulatory functions $g_\obu(\sv)$, $g_\obp(\sv)$, $g_\ocu(\sv)$ and $g_\ocp(\sv)$ in Eqs~\eqref{obp}--\eqref{oca} include a geometrical feedback at various stages of osteoblast and osteoclast development. This enables us to distinguish two types of geometrical action. Indeed, the modulation of the bone cell developmental stages by the specific surface can be interpreted in the model in terms of modulation of the initiation of new remodelling events (\bmu\ creation) and modulation of resorption and formation within existing \bmu s, as explained in the following:

\begin{enumerate}
\item \emph{Initiation of new remodelling events.} The initiation of a new remodelling event is a localised process that creates a new \bmu. While the exact biochemical mechanisms that lead to the creation of a new \bmu\ are poorly understood, it is believed that first steps in this process are the recruitment of pre-osteoclasts and pre-osteoblasts at the bone surface~\cite{martin-burr-sharkey}. This recruitment is thought to be controlled by osteocytes sensing the local mechanical state of the bone matrix and communicating with progenitor cells in the marrow through the bone surface. The complex dependence of bone surface availability on the recruitment of pre-osteoblasts and pre-osteoclasts is modelled by the geometrical regulation of cell differentiation exerted by the functions $g_\obu(\sv)$ and $g_\ocu(\sv)$. Therefore, the geometrical regulation by $g_\obu(\sv)$ and $g_\ocu(\sv)$ models the influence of bone surface availability on the initiation of new remodelling events, and so on the number of \bmu s in the representative volume element, which in turn determines the bone turnover rate~\cite{parfitt2-in-recker}. This type of geometrical regulation of bone remodelling is similar to the geometrical regulation of the `activation frequency' of \bmu s used by Hazelwood~\etal~\cite{hazelwood-etal}.

\item \emph{Modulation of resorption and formation within existing \bmu s.} Active osteoclasts can only resorb bone from the bone surface. Similarly, active osteoblasts are only observed to deposit new bone at the bone surface. The maturation of pre-osteoblasts and pre-osteoclasts into active cells thus depends on bone surface availability. This dependence is modelled by the geometrical regulation of cell differentiation exerted by the functions $g_\obp(\sv)$ and $g_\ocp(\sv)$. Therefore, these functions dictate how many active osteoclasts and active osteoblasts can form in \bmu s that are already remodelling bone (\ie, which contain already pre-osteoblasts and pre-osteoclasts), and so how much bone is resorbed and formed in existing \bmu s, which in turn determines bone balance. Note that while bone surface availability is necessary for cell activation, it is not sufficient. For instance, osteoclasts can only become active if, in addition, their receptor \rank\ is activated by the ligand \rankl.
\end{enumerate}

Due to the different ways of action of bone surface on cell differentiation explained above, the regulatory functions $g_\obu(\sv)$, $g_\obp(\sv)$, $g_\ocu(\sv)$ and $g_\ocp(\sv)$ may take different forms and can be complicated functions of the specific surface $\sv$. We take here a phenomenological approach and assume that each of these functions can be represented by a power law of \sv:
\begin{align}
    g_i(\sv) = \Big(\frac{\sv}{\sv(t_0)}\Big)^{k_i}, \quad \text{with}\ k_i \geq 0, \quad i=\obu, \obp, \ocu, \ocp.\label{g-definition}
\end{align}
In Eq.~\eqref{g-definition}, $\sv(t_0)$ denotes the specific surface in the bone remodelling steady state, which is assumed to be homeostatic (no bone gain or loss, but a steady bone turnover). The normalisation of \sv\ by its steady-state value $\sv(t_0)$ ensures that in the steady state, $g_i=1\ \forall i$, and so that the steady state of the model is consistent with previous models of bone remodelling without geometrical regulation~\cite{pivonka-etal-1,scheiner-etal-mechanostat}. The benefit of using a power-law function of \sv\ in Eq.~\eqref{g-definition} is that geometrical feedback can be ``switched off'' by choosing $k_i=0\ \forall i$ in Eq.~\eqref{g-definition}. In this situation, not only the steady state, but also the dynamical behaviour of the model of Ref.~\cite{scheiner-etal-mechanostat} (including mechanical feedback) is retrieved. Mechanical feedback is ``switched off'' by choosing $\kappa=0$, $\lambda=0$ in Eqs~\eqref{Prankl} and~\eqref{Pieps}. Note that when both geometrical and mechanical feedbacks are ``switched off'', the bone cell population model of Refs~\cite{pivonka-etal-1,pivonka-etal-2} is retrieved, except for a remaining pre-osteoblast proliferation term in Eq.~\eqref{obp} that was not accounted for previously.\footnote{Compared to Refs~\cite{pivonka-etal-1,pivonka-etal-2}, the differentiation rate of \obu s to \obp s is reduced accordingly to ensure that the model converges to the same steady state.}

To reveal in which ways the morphology of the microstructure of bone may influence bone remodelling, we will investigate the effects of several combinations of $g_\obu$, $g_\obp$, $g_\ocu$ and $g_\ocp$ in Section~\ref{sec:results} and determine combinations that lead to physiologically meaningful results.

\subsection{Changes in porosity and bone matrix fraction due to cell activity}
The activity of osteoclasts and osteoblasts leads to the removal and deposition of new bone. This activity modifies the volume fraction of bone matrix in the tissue. Osteoblasts deposit osteoid, a collagen-rich substance which later mineralises into new bone. Primary mineralisation of osteoid is relatively fast: 70\% of the maximum mineral density is reached within a few days in humans~\cite{parfitt2-in-recker}. Given the much larger time spans involved in the remodelling processes, it is fair to model the osteoblasts ``instantaneously'' depositing ``fully'' mineralised new bone matrix, as was assumed in Refs~\cite{pivonka-etal-1,pivonka-etal-2,scheiner-etal-mechanostat}. We further assume that the resorption rate of bone matrix $\kres$ by an individual active osteoclast (in volume per unit time) is constant, and that the rate of new bone matrix deposition $\kform$ by an individual osteoblast (in volume per unit time) is constant. The evolution of the vascular porosity and bone matrix volume fraction are thus given by
\begin{align}
    &\frac{\d}{\d t}\fvas = -\frac{\d}{\d t}\fbm = - \kform \oba + \kres\oca
    \label{fvas}.
\end{align}

\subsection{Comparison with the model of geometrical regulation of bone remodelling by Martin}
It is informative at this point to compare our formulation of geometrical regulation of bone remodelling with that proposed by Martin in Ref.~\cite{martin-1984}. The evolution of the vascular porosity proposed by Martin is, in our notations (see~\cite[Eq.~(67)]{martin-1984}):\footnote{In Ref.~\cite{martin-1984}, Martin uses the symbol $S_\lambda$ to denote only the fraction of 3D specific surface that is capable of remodelling. This excludes surfaces of the lacunae-canaliculi system~\cite{martin-1972}. In this paper, we do not consider surfaces of the lacunae-canaliculi system, and so $S_\lambda$ in Ref.~\cite{martin-1984} is equal to our definition of $\sv$.}
\begin{align}
    \frac{\d}{\d t}f_\text{vas, Martin} = -(\kform\, \delta_\oba\, \lambda_\oba - \kres\,\delta_\oca\,\lambda_\oca)\ \sv
    \label{fvas-martin}
\end{align}
where $\delta_\oba$ and $\delta_\oca$ are the active osteoblast and active osteoclast surface densities at an active bone site (number of cells per unit surface), and $\lambda_\oba$ and $\lambda_\oca$ are the fractions of the total available bone surface in which there is osteoblastic and osteoclastic activity.  The quantity $\kform\, \delta_\oba\, \lambda_\oba \sv$ is the formation rate of bone matrix and corresponds to $\kform \oba$ in our model. The quantity $\kres\, \delta_\oca\, \lambda_\oca \sv$ is the resorption rate of bone matrix and corresponds to $\kres \oca$ in our model. To model an osteoporotic pathological condition, Martin assumes a constant bone imbalance between resorption and formation in remodelling \bmu s. This imbalance is modelled by setting $(\kform\,\delta_\oba\,\lambda_\oba - \kres\,\delta_\oca\,\lambda_\oca) = -2$ \um/year in Eq.~\eqref{fvas-martin}, meaning that in average, a 2 \um-thick layer of the bone surface is resorbed each year. The results presented in Figure~\ref{fig:martin} were obtained by Martin in Ref.~\cite{martin-1984} by integrating Eq.~\eqref{fvas-martin} with this constant imbalance and with either $\sv\equiv 1$ (no geometrical feedback) or using the function $\sv(\fvas)$ presented in Eq.~\eqref{sv-fvas} to include the effect of geometrical feedback.

A limitation of the formulation by Martin is that the dependence of the rate of change of porosity upon \sv\ does not account for microscopic effects of bone surface availability on the recruitment and development of bone cells. Indeed, in Martin's formulation, the complexity of osteoblast and osteoclast recruitment and development at a remodelling bone surface is embodied by the fractions $\lambda_\oba$ and $\lambda_\oca$, but these fractions are assumed to be independent of $\sv$. Martin and others have interpreted Eq.~\eqref{fvas-martin} as the influence of geometrical regulation on the activation frequency of \bmu s~\cite{martin-1984,parfitt2-in-recker}. Indeed, the densities of active osteoblasts and osteoclasts in Martin's model are proportional to the specific surface. Comparing Eqs~\eqref{fvas} and~\eqref{fvas-martin}, one sees that  $\oba = \lambda_\oba \sv \delta_\oba$ and $\oca = \lambda_\oca \sv \delta_\oca$. Thus, an increased specific surface represents increased numbers of remodelling \oba s and \oca s, and so an increased number of \bmu s. 

The strength of our model is to consider several stages in the recruitment and development of osteoblasts and osteoclasts explicitly. This enables us to include a more detailed geometrical feedback acting directly on these different stages of the remodelling sequence. In our model, even steady-state values of \oba s and \oca s are complex functions of the specific surface. This complexity represent the implicit dependence of $\lambda_\oba$ and $\lambda_\oca$ upon \sv\ not accounted for by Martin. Additionally, no mechanical feedback has been considered by Martin. In Ref.~\cite{scheiner-etal-mechanostat}, Scheiner \etal\ have discussed the problem of neglecting mechanical feedback in models of bone remodelling applied to catabolic bone diseases (such as osteoporosis), which leads to a unlimited bone loss which is physiologically not observed. This behaviour can be seen in Figure~\ref{fig:martin} for cortical bone where the vascular porosity continues to increase (both with and without geometrical regulation). It is known from bone mineral density measurements in osteoporotic patients that the increase in bone porosity in osteoporosis slows down with time. The porosity eventually reaches an upper bound whose value depends on the patient.  We note here that the long-term value of the vascular porosity $\fvas$ reached with mechanical feedback in our model depends in particular on the initial value of \fvas\ (see Section~\ref{sec:results}).

\section{Numerical Results and discussion}\label{sec:results}
Having incorporated biochemical, biomechanical and geometrical regulation of bone remodelling we are now in a position to investigate the effects of various regulatory parameters on changes in bone cell numbers, vascular porosity and bone stiffness. To compare our model with the model suggested by Martin in Ref.~\cite{martin-1984}, we simulate an underlying osteoporotic condition as a perturbation from the original (homeostatic) steady state situation. 

Osteoporosis is a bone disease that leads to an increase in porosity in both cortical and trabecular bone~\cite{nishiyama-etal}. This increase in porosity generates a progressive reduction in bone stiffness and a higher fracture risk. To simulate an osteoporotic condition in our model, we perturb the homeostatic steady state (a state with no bone loss or gain and a constant bone turnover rate) by increasing the \rankl/\opg\ ratio. This can be achieved in our model by prescribing an excess of $\pth$ concentration (see Refs.~\cite{lemaire-etal,pivonka-etal-1}).\footnote{The physiological effect of \pth\ administration for bone remodelling is complex and in particular, depends on the time course of the administration~\cite{frolik-etal}. Continuous \pth\ administration (infusion) leads to bone loss with increased turnover. However, intermittent \pth\ administration (daily injections) leads to bone gain~\cite{frolik-etal,burr-etal}. Only the `continuous' action of \pth\ is represented in our model~\cite{lemaire-etal,pivonka-etal-1}. We have shown previously that increasing \pth\ leads to an increase in the \rankl/\opg\ ratio, and so to bone loss with higher turnover rate compared to the homeostatic bone remodelling state~\cite{pivonka-etal-2,scheiner-etal-denosumab}. Increasing \pth\ concentration thus consists in an adequate representation of osteoporosis in the present model. We also note that the effect of \pth\ in bone remodelling is known to interact synergistically with mechanical loading~\cite{ma-etal,turner-etal}. Consistently with the interpretation of increasing \pth\ as a representation of osteoporosis in our model, we do not consider this synergistic interaction here.} The osteoporotic condition is assumed to develop instantly from an initial homeostatic state at time $t_0$ (corresponding to a middle biological age). To obtain a steady increase in $\fvas$ of 0.01/year without geometrical regulation, as has been assumed by Martin (see Figure~\ref{fig:martin}), a continuous \pth\ administration rate $P_\pth(t) \equiv 500~\pM/\da$ has been applied at all times $t>t_0$. We denote by $\Delta t_\text{OP} = t-t_0$ the time elapsed since the onset of the osteoporotic condition. The evolution of the system is followed for a period of time of 20 years from the onset of osteoporosis, \ie, $0 \leq \Delta t_\text{OP} \leq \text{20 years}$. The initial values for vascular porosity of cortical and trabecular bone have been chosen as $f_\vas^\text{cort.} = 0.05$ and $f_\vas^\text{trab.} = 0.75$. For the simulations including biomechanical regulation, a uniaxial compressive stress $\boldsymbol\Sigma=\Sigma^\text{uni} \textbf{e}_3\otimes\textbf{e}_3$ with $\Sigma^\text{uni}=- 30~\text{MPa}$ has been assumed to exert on the representative volume element.

\subsection{Simulation of osteoporosis: Evolution of bone porosity and bone stiffness properties}
\begin{figure}[h!]
    \centering\includegraphics[width=\figurewidth]{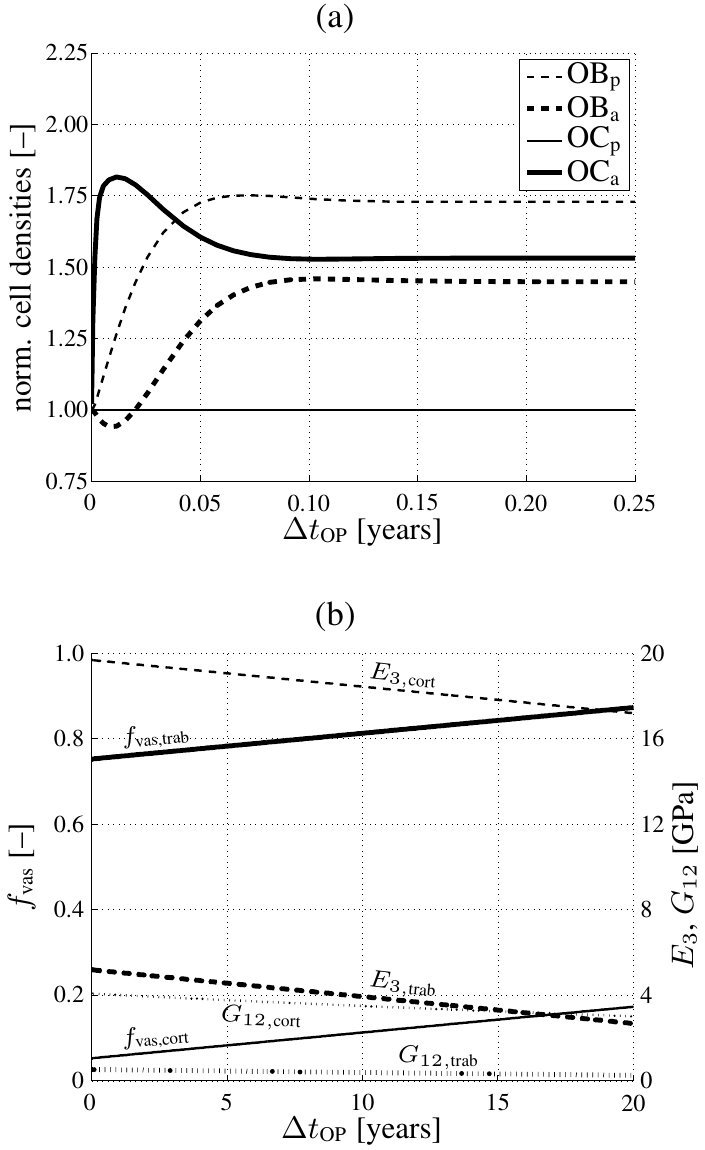}%
    \caption{Time evolutions for a simulated osteoporotic condition of (a) bone cell densities (normalised to the initial homeostatic steady state) and (b) the vascular porosities \fvas, axial stiffnesses $E_3$ and shear modulus $G_{12}$ in both a cortical bone tissue sample and a trabecular bone tissue sample under uniaxial compression. In this figure, no geometrical nor mechanical feedback is considered.}
    \label{fig:time-evolutions}
\end{figure}
Figure~\ref{fig:time-evolutions} shows the simulated evolution of bone cell densities, vascular porosity ($\fvas$) and selected components of the bone stiffness matrix (\ie, axial stiffness $E_3$ and shear modulus $G_{12}$) in osteoporosis considering no geometrical feedback and no mechanical feedback. The evolution of bone cell densities exhibits a short transient for a period of $\approx30$~days after the onset of osteoporosis (Figure~\ref{fig:time-evolutions}a). After this initial transient, \oca s and \oba s reach a new steady state, in which resorption and formation are imbalanced and lead to osteoporotic bone loss. Note that a short time interval has been chosen in Figure~\ref{fig:time-evolutions}a to demonstrate the short transient cellular response, while a larger time interval of 20~years is chosen to follow the evolution of the vascular porosity, axial stiffness and shear modulus. The progressive increase in vascular porosity is shown in Figure~\ref{fig:time-evolutions}b together with the associated reduction in bone stiffness. After 20~years of osteoporosis, the normal stiffness in the longitudinal direction of cortical bone, $E_3$, has dropped by about 20\%, whereas the shear modulus of cortical bone,  $G_{12}$, has dropped by about 40\%. From these results it is clear that the evolution of vascular porosity drives the changes in bone stiffness differently for the different components of the stiffness tensor~\cite{scheiner-etal-denosumab}. Generally speaking, an increase in \fvas\ is always associated with a reduction of bone stiffness properties. For conciseness, in the following we will only present the effects of geometrical and biomechanical regulation on the vascular porosity \fvas\ alone.

\subsection{Geometrical regulation---effect on individual stages of osteoblast and osteoclast developments}
\begin{figure*}[h!]
    \centering\includegraphics[width=0.9\textwidth]{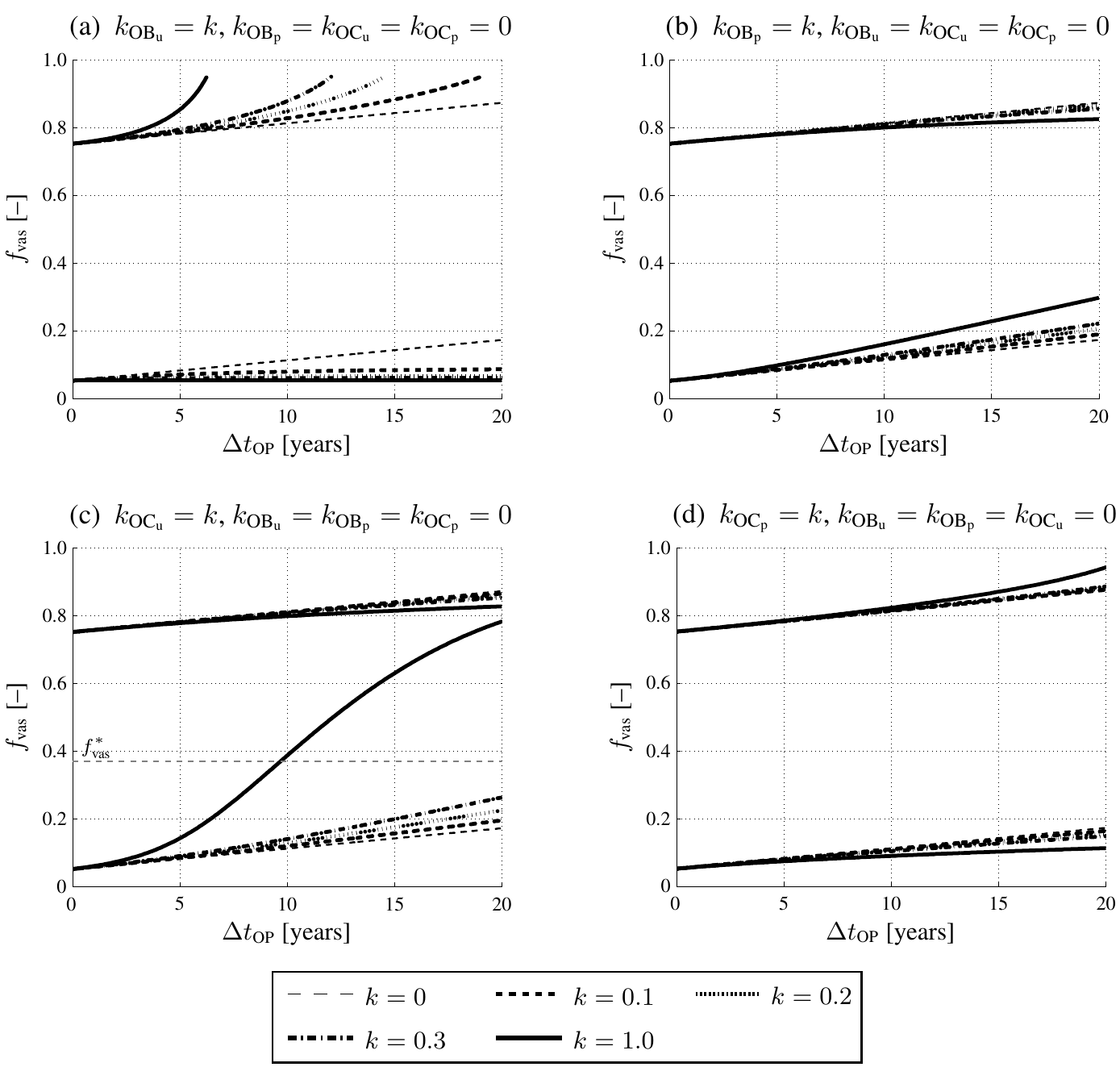}
    \caption{Influence of geometrical regulation on the increase of vascular porosity in osteoporosis. In this figure, geometrical regulation is included at a single stage of osteoblast or osteoclast development only. The numerical simulations are initiated from either cortical bone (lower curves) or trabecular bone (upper curves). The effect of several strengths $k$ of the geometrical regulations are shown (corresponding to the exponent of the power law in Eq.~\eqref{g-definition}). The case $k=0$ corresponds to no geometrical feedback. (a) Geometrical regulation of \obu\ differentiation only; (b) geometrical regulation of \obp\ differentiation only; (c) geometrical regulation of \ocu\ differentiation only; (d) geometrical regulation of \ocp\ differentiation only.}
    \label{fig:geometrical-regulation-single}
\end{figure*}
We first present how geometrical regulation influences the osteoporotic increase in bone porosity \fvas\ without accounting for mechanical feedback, \ie, by setting $\kappa = 0~\pM/\da$ and $\lambda = 0$ in Eqs~\eqref{Prankl}, \eqref{Pieps}.
Based on the four regulatory functions $g_\obu$, $g_\obp$, $g_\ocu$, and $g_\ocp$ accounting for geometrical regulation, we first performed $2^4=16$~simulations `switching on' or `off' each of these regulatory functions to investigate all different possible combinations. However, it turns out that among all possible simulations there are only three patterns which reflect the bone systems behaviour. These patterns can best be studied by looking first at the cases in which only a single regulatory function is `switched on' while all others are `switched off' (\ie, identically set to one) (Figure~\ref{fig:geometrical-regulation-single}). We will present a selection of combinations of geometrical regulation on several stages of osteoblast and osteoclast development in Figures~\ref{fig:geometrical-regulation-combined-first} and~\ref{fig:geometrical-regulation-combined-second}.

Figure~\ref{fig:geometrical-regulation-single}a shows the influence of $g_\obu(\sv)$ on the increase of vascular porosity in osteoporosis compared to the case of no geometrical and no biomechanical regulations in both cortical and trabecular bone. The time evolution of the porosity $\fvas(t)$ in Figure~\ref{fig:geometrical-regulation-single}a clearly shows that while $g_\obu(\sv)$ reduces bone loss in cortical bone it accelerates bone loss in trabecular bone. The strength of this regulation depends on the exponent $k_\obu$ of the normalised geometrical regulatory function in Eq.~\eqref{g-definition}. Exponents in the range $0.3 \lesssim k_\obu \lesssim 1$ exhibit a relatively strong regulatory effect on the evolution of the disease, while exponents in the range $0 <k_\obu \lesssim 0.2$ only exhibit a moderate effect. Interestingly, the mechanism of action of $g_\obu$ is opposite to the geometrical regulation obtained by Martin~\cite{martin-1984} (see Figure~\ref{fig:martin}). In cortical bone, the osteoporotic increase in \fvas\ induces an increase in \sv\ (see Figure~\ref{sv-fvas}), and so an increase of $g_\obu(\sv)$. This increases in turn the generation of osteoblasts, which has an anabolic (\ie, bone forming) effect and stabilises the osteoporotic loss of bone~\cite{lemaire-etal,pivonka-etal-1}. By contrast, in trabecular bone the osteoporotic increase in \fvas\ induces a decrease in \sv\ (see Figure~\ref{sv-fvas}), and so a decrease of $g_\obu(\sv)$. This decreases in turn the generation of osteoblasts, which accelerates the osteoporotic loss of bone.

While this scenario can be understood from a theoretical point of view, it has to be emphasised that it is probably physiologically unrealistic. As argued in Section~\ref{sec:geom-regulation}, the geometrical regulation of \obu\ differentiation is associated with the creation of a new remodelling event, \ie, of a new \bmu. However, both \obu\ differentiation and \ocu\ differentiation are activated in such an event. Here, the consideration of a geometric regulation on \obu\ differentiation alone represents the initiation of a formation event and is not associated with a joint initiation of a resorption event. To represent the geometrical regulation of \bmu\ creation, both $g_\obu(\sv)$ and $g_\ocu(\sv)$ should act together (see Figures~\ref{fig:geometrical-regulation-combined-first} and~\ref{fig:geometrical-regulation-combined-second}).

Figure~\ref{fig:geometrical-regulation-single}b shows the influence of $g_\obp(\sv)$ on the increase of \fvas\ in osteoporosis. In this figure, \fvas\ is only weakly affected by the geometrical feedback in both cortical and trabecular bone. The effect of bone surface availability on the differentiation of pre-osteoblasts into active osteoblasts modelled by $g_\obp(\sv)$ does not seem to affect the bone remodelling balance significantly in the first 10~years. In fact, neither the population of active osteoblasts nor the population of active osteoclasts is affected significantly, while the population of pre-osteoblasts is decreased in cortical bone and increased in trabecular bone (not shown). The increased differentiation rate of pre-osteoblasts into active osteoblasts in cortical bone (due to the increase in the available bone surface) depletes the population of pre-osteoblasts. The population of active osteoblasts is thus derived from a smaller pool of pre-osteoblasts that are differentiating faster, and therefore stays relatively constant. The opposite effect occurs in trabecular bone, \ie, a larger pool of pre-osteoblasts is created but they are differentiating into active osteoblasts more slowly.

Figure~\ref{fig:geometrical-regulation-single}c shows the influence of $g_\ocu(\sv)$ on the increase of \fvas\ in osteoporosis. The behaviour of \fvas\ obtained for this type of geometrical regulation is in general agreement with the geometrical regulation obtained by Martin~\cite{martin-1984} (see Figure~\ref{fig:martin}), \ie, geometrical regulation leads to increased bone loss in cortical bone and decreased bone loss in trabecular bone. In cortical bone, the increase in $g_\ocu(\sv)$ due to the osteoporotic condition leads to an increase in osteoclastogenesis, which by biochemical coupling also increases (although to a lesser extent) the population of osteoblasts. This results in a high turnover rate with a catabolic bias, \ie, a high rate of bone loss. In trabecular bone, the situation is reversed since $g_\ocu(\sv)$ decreases with the evolution of the osteoporotic condition. A state with lower turnover rate and a reduction in resorption is reached. The low turnover rate explains the reduced amplitude of the response in trabecular bone compared to that seen in cortical bone.

The strength of the geometrical regulation is determined by the value of $k_\ocu$ and is seen to strongly affect bone balance. In particular for the case of strong modulation, \ie, $k_\ocu=1$, a rapid acceleration of cortical bone loss occurs over the first 10~years of osteoporosis. At nearly 10~years, a transition is taking place leading to a reduction in the rate of bone loss. This behaviour is due to the fact that at this time, the bone porosity reaches the critical value $\fvas^\ast\approx 0.37$ at which the specific surface is maximum (see Figure~\ref{fig:sv-fvas}). In the first 10~years of osteoporosis, $\fvas$ is in the ascending branch of $\sv(\fvas)$ and so $g_\ocu(\sv)$ increases, while after reaching the maximum specific surface, $g_\ocu(\sv)$ decreases, which leads to a reduction in the rate of bone loss. Over 20~years, the original cortical bone has been resorbed enough to reach trabecular porosities. Such a strong effect of bone remodelling on bone porosity is normally not physiologically seen. However, it is possible that in osteoporosis, a locally strong geometrical feedback may help initiate or accentuate the observed `trabecularisation' of bone, by which cortical bone is progressively lost and transformed into trabecular bone at the endocortical wall, leading to cortical wall thinning and expansion of the medullary cavity~\cite{mayhew-clement-etal,thomas-feik-clement,cooper-etal-2007,seeman}. Indeed, in bone, the endocortical wall exhibits the highest specific surface and is known to be highly remodelling. It can be expected that geometrical regulation plays a particularly significant role in this region of bone.

Figure~\ref{fig:geometrical-regulation-single}d shows the influence of $g_\ocp(\sv)$ on the increase of \fvas\ in osteoporosis. This influence is qualitatively similar to that of $g_\ocp$ shown in Figure~\ref{fig:geometrical-regulation-single}a, \ie, a reduction in bone loss in cortical bone and an increase in bone loss in trabecular bone. However, the geometrical regulation modelled by $g_\ocp$ is less pronounced than the geometrical regulation modelled by~$g_\obu$.

\subsection{Geometrical regulation---combined effect on several stages of osteoblast and osteoclast developments}
\begin{figure}[t!]
    \centering%
    \includegraphics[width=0.85\figurewidth]{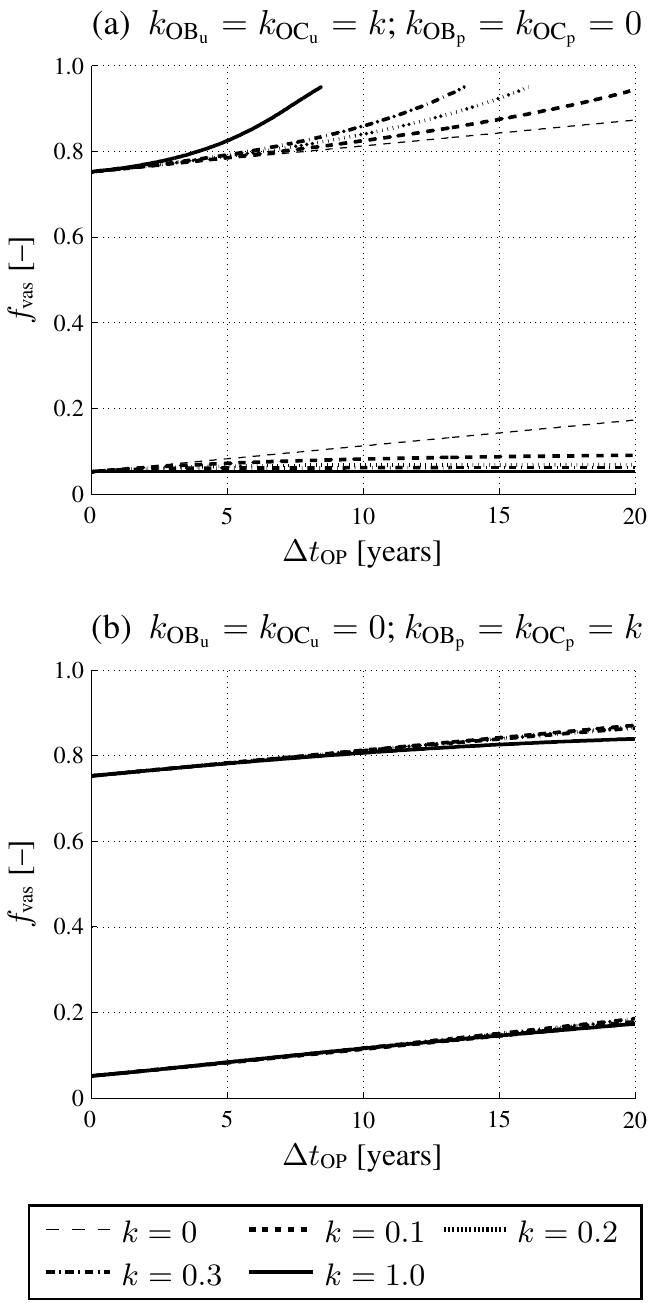}
    \caption{Combined influence of several geometrical regulations on the increase of vascular porosity in osteoporosis. The numerical simulations are initiated from either cortical bone (lower curves) or trabecular bone (upper curves). The effect of several strengths $k$ of the geometrical regulations are shown (corresponding to the exponent of the power law in Eq.~\eqref{g-definition}). The case $k=0$ corresponds to no geometrical feedback. (a) Joint geometrical regulation of both \obu\ differentiation and \ocu\ differentiation; (b) joint geometrical regulation of both \obp\ differentiation and \ocp\ differentiation.}
    \label{fig:geometrical-regulation-combined-first}
\end{figure}

As mentioned previously, a geometrical regulation of the creation of new remodelling events (\ie\ new \bmu s) should involve a regulation of both the recruitment of osteoclasts and the recruitment of osteoblasts (see Section~\ref{sec:geom-regulation}). In Figure~\ref{fig:geometrical-regulation-combined-first}a, we show the combined influence of both $g_\ocu(\sv)$ and $g_\obu(\sv)$ on the increase of \fvas\ in osteoporosis. The exponents $k_\ocu$ and $k_\obu$ measuring the strength of the geometrical regulation are assumed identical. By comparing with the individual influences of $g_\obu(\sv)$ and $g_\ocu(\sv)$ in Figure~\ref{fig:geometrical-regulation-combined-first}a and~\ref{fig:geometrical-regulation-combined-first}c, one sees that the geometrical regulation of osteoblast recruitment overrides that of osteoclast recruitment. As a consequence, the overall behaviour is opposite to that obtained by Martin~\cite{martin-1984} for the geometrical regulation of the creation of new \bmu s (see Figure~\ref{fig:martin}).

By contrast, in Figure~\ref{fig:geometrical-regulation-combined-first}b, the geometrical regulation of the last stage of osteoclast differentiation by $g_\ocp(\sv)$ and the geometrical regulation by $g_\obp(\sv)$ of the last stage of osteoblast differentiation (modelling a regulation of the activation of cells within existing \bmu s) seem to compensate each other and to result in an evolution of the porosity that is almost unaffected by geometrical feedback. The geometrical regulations assumed in Figures~\ref{fig:geometrical-regulation-combined-first}a and~\ref{fig:geometrical-regulation-combined-first}b represent different natures of the influence of bone surface availability on bone remodelling (see Section~\ref{sec:model}). Therefore, our simulations suggest that for the simulated evolution of bone porosity in osteoporosis, the influence of surface availability is significantly stronger on the creation of new remodelling events (new \bmu s) (Figure~\ref{fig:geometrical-regulation-combined-first}a) than on the activation of bone cells within already active remodelling sites (within existing \bmu s).

\begin{figure}[t!]
    \centering%
    \includegraphics[width=0.9\figurewidth]{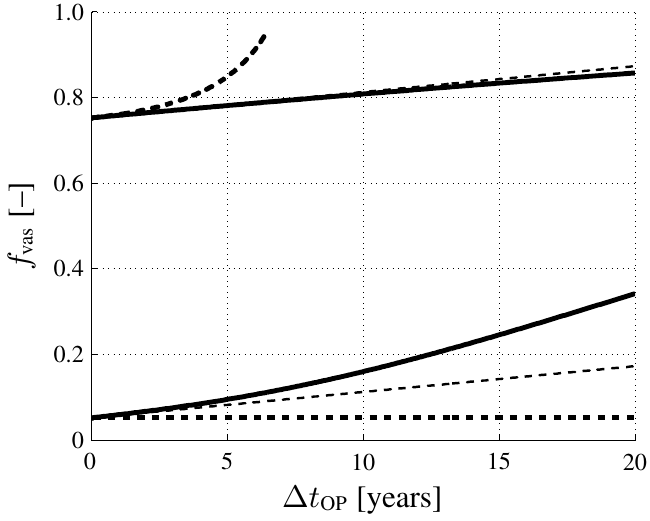}
    \caption{Influence of the geometrical regulation of \bmu\ creation on the increase of vascular porosity in osteoporosis. Different strengths for the geometrical regulation of \obu\ differentiation and \ocu\ differentiation are investigated via the exponents $k_\obu$ and $k_\ocu$ of the regulatory functions $g_\obu(\sv)$ and $g_\ocu(\sv)$: $k_\obu=0, k_\ocu=0$ (no geometrical regulation, thin dashed lines); $k_\obu=1, k_\ocu=0.1$ (thick dashed lines) and $k_\obu=0.1, k_\ocu=1$ (thick solid line).}
    \label{fig:geometrical-regulation-combined-second}
\end{figure}

In Figure~\ref{fig:geometrical-regulation-combined-second}, we show that a similar influence of geometrical regulation of \bmu\ creation as that obtained by Martin~\cite{martin-1984} can be retrieved in our model by modifying the relative strengths of the regulatory functions $g_\obu(\sv)$ and $g_\ocu(\sv)$ via the exponents $k_\obu$ and $k_\ocu$. The vascular porosity can exhibit a wide range of behaviours, interpolating between the situation of Figure~\ref{fig:geometrical-regulation-single}a ($k_\obu = 1$, $k_\ocu=0$) and the situation of Figure~\ref{fig:geometrical-regulation-single}c ($k_\obu=0$, $k_\ocu=1$). We conclude that geometrical feedback has the potential to significantly influence the evolution of bone diseases. However, complex biochemical coupling between osteoblasts and osteoclasts makes it difficult to predict the relative strength of the influence of geometrical regulation on osteoblast and osteoclast developments.

Finally, we note that because the curve $\sv(\fvas)$ exhibits a maximum at $\fvas^\ast\approx 0.37$ (see Figure~\ref{fig:sv-fvas}), geometrical feedback always induces opposite behaviours for the evolution of vascular porosity in cortical bone (for which $\fvas < \fvas^\ast$) and in trabecular bone (for which $\fvas > \fvas^\ast$).

\subsection{Coupled geometrical and mechanical regulations}
\begin{figure}[t]
    \centering%
    \includegraphics[width=0.9\figurewidth]{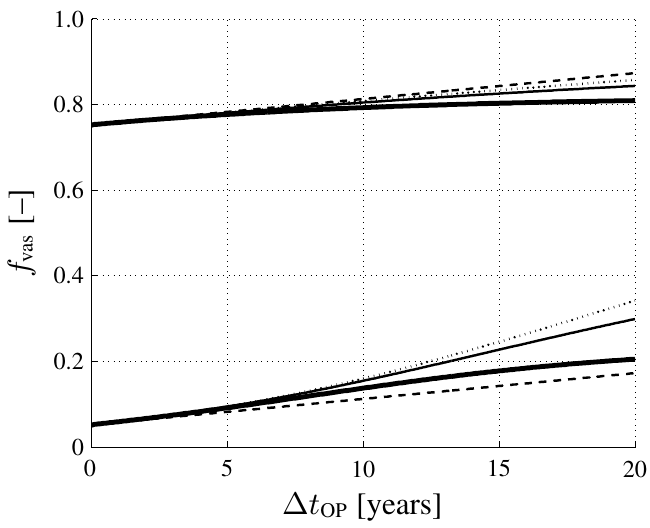}
    \caption{Influence of coupled geometrical and mechanical regulation of bone remodelling on the increase of vascular porosity in osteoporosis. The evolution of \fvas\ without geometrical regulation and without mechanical regulation is shown as dashed lines. The inclusion of geometrical regulation without mechanical regulation is shown as dotted lines. This geometrical regulation is assumed identical to the solid line in Figure~\ref{fig:geometrical-regulation-combined-second}, \ie, $k_\obu = 0.1$, $k_\ocu = 1$, $k_\obp=k_\ocp=0$. Two cases of coupled geometrical and mechanical regulations are shown as solid lines for two strengths $\lambda$ of the biomechanical regulation: $\lambda=0.1$ (thin solid lines) and $\lambda=0.5$ (thick solid lines).}
        \label{fig:geometrical-mechanical-regulation}
\end{figure}

Figure~\ref{fig:geometrical-mechanical-regulation} shows the effect of adding a mechanical regulation of bone remodelling for the evolution of the vascular porosity. The geometrical regulation considered in Figure~\ref{fig:geometrical-mechanical-regulation} (see dotted line) is assumed to represent the influence of bone surface availability for \bmu\ creation as in Figure~\ref{fig:geometrical-regulation-combined-second}, solid line. Two strenghts of the biomechanical transduction are illustrated (\ie, $\lambda=0.1$ and $\lambda=0.5$ in Eq.~\eqref{Pieps}). Figure~\ref{fig:geometrical-mechanical-regulation} suggests that mechanical feedback has the potential to progressively override both the evolution of osteoporosis and the influence of geometrical regulation. Indeed, mechanical feedback is seen to stabilise bone balance, irrespective of whether geometrical feedback stabilises or destabilises bone balance in osteoporosis. Such a stabilisation of bone loss is clinically observed in osteoporotic patients.

Similarly to the simulations by Scheiner \etal~\cite{scheiner-etal-mechanostat}, mechanical feedback counteracts bone loss in osteoporosis both in cortical and trabecular bone, on a time scale that depends on the strength of mechanical regulation (\ie, on the parameter $\lambda$). Interestingly, in our simulations with geometrical feedback, the strength $\lambda$ of the mechanical regulation also has an influence on the steady-state value of the porosity attained. This is to be contrasted to the situation without geometrical feedback in which no such influence was observed~\cite{scheiner-etal-mechanostat}. In fact, without geometrical feedback, the vascular porosity \fvas\ enters the right hand side of the governing equations~\eqref{obp}--\eqref{oca}, \eqref{fvas} only implicitly via the strain energy density $\sed$, see Eq.~\eqref{sed}. Consequently, the steady-state value of \fvas\ is uniquely determined by the steady-state value of the strain energy density, $\sedst$, and the macroscopic loading $\boldsymbol\Sigma$, which are themselves independent of the strength $\lambda$ of biomechanical regulation~\cite{scheiner-etal-mechanostat}. With geometrical feedback, additional dependences upon the vascular porosity \fvas\ are added in the governing equations for the bone cells, Eqs~\eqref{obp}--\eqref{oca}. Consequently, the steady-state value of \fvas\ now depends in addition on the biochemical and cellular state of the system, which depends in turn on the strength $\lambda$ of the biomechanical regulation of the bone cells.

From the above discussion, it is clear that the coupling between the geometrical and mechanical feedbacks is effectively mediated by the biochemical and cellular state of the system. In osteoporotic patients, both a geometrical and a mechanical feedback are present at the same time, as well as biochemical and hormonal dysregulations underlying the establishment of osteoporosis. The interdependence between the biochemical and hormonal dysregulations in osteoporosis, geometrical feedback and mechanical feedback, is not trivial to elucidate. Physiologically, it is expected that all these influences play a role for the evolution of bone vascular porosity. Our mathematical model is a first attempt to integrate these influences and help understand their contribution for clinically observed changes of bone porosities in osteoporotic patients.

\section{Conclusions}
In this paper we developed a computational model of bone remodelling that takes into account biochemical, biomechanical and geometrical regulations of bone cells. The biochemical regulation of the bone cells is based on the model developed in Refs~\cite{pivonka-etal-1,pivonka-etal-2}. The biomechanical regulation of the bone cells is based on the model developed in Ref.~\cite{scheiner-etal-mechanostat}, in which a continuum micromechanical approach is used to consistently link bone cell responses with mechanical properties of bone. The new contribution of this paper is the inclusion of a geometrical regulation of bone cells in the model. Geometrical feedbacks were included at several developmental stages of osteoblasts and osteoclasts to represent the influence of microscopic bone surface availability for various bone microstructures. We investigated the influence of a geometrical regulation of bone cells in bone remodelling both without and with consideration of biomechanical regulation for a simulated osteoporotic condition. From the numerical simulations, we identified the following actions of geometrical and mechanical regulation on bone remodelling:
\begin{itemize}
    \item Geometrical regulation of bone remodelling may play an important role for the initiation of new \bmu s as described by the combined effect of the geometrical regulatory functions $g_\obu$ and $g_\ocu$ acting on \obu\ differentiation and \ocu\ differentiation; in particular geometrical regulation of \ocu s via $g_\ocu$ seems to be most important to retrieve similar evolutions of bone porosities in osteoporosis as obtained by Martin~\cite{martin-1984};
    \item Geometrical regulation of \bmu\ creation affects cortical and trabecular bone in opposite ways in osteoporosis: while bone resorption is enhanced in cortical bone due to the increase in specific surface with increasing porosity, bone resorption is slowed down in trabecular bone due to the decrease in specific surface with increasing porosity;
    \item Geometrical regulation of the activation of osteoblasts and osteoclasts in existing \bmu s seems to play a secondary role for the evolution of osteoporosis. While the specific surface can influence the differentiation of bone precursor cells into active resorbing/forming cells, no significant influence on bone porosity was observed in our simulations.
    \item Our simulations suggest that geometrical regulation may play a role in osteoporosis for the initiation and\slash or accentuation of the observed `trabecularisation' of bone at the endocortical wall. At the endocortical wall, the specific surface and bone turnover are high and so the effects of a geometrical feedback can be expected to be significant;
    \item Our simulations of coupled geometrical and mechanical regulations suggest that the stabilisation of bone loss observed clinically in osteoporotic patients is probably accelerated by geometrical feedback in trabecular bone, but is probably slowed down by geometrical feedback in cortical bone.
    \item Both mechanical and geometrical feedbacks are important to account for in our model of bone remodelling. Mechanical feedback enables the local porosity of bone tissue $\fvas$ to stabilise to a well-defined value within $[0,1]$. Geometrical feedback enables this value to be not only determined by the external loading of the tissue, but also by the biochemical and cellular state of the system, as would be expected physiologically.
\end{itemize}

\section*{Acknowledgements}
Financial support by the Australian Research Council (ARC), in the framework of the project \emph{Multiscale modelling of transport through deformable porous materials} (project number DP-0988427) and by the European Research Council (ERC) in the framework of the project \emph{Multiscale poro-micromechanics of bone materials, with links to biology and medecine} (project number FP7-257023) are gratefully acknowledged.

\begin{appendices}
\section{Model description}\label{appx:model-description}
In this appendix, we complete the mathematical description of the model and give the values of the parameters. The nomenclature used in the paper is split into three tables. Table~\ref{table:dynamic-variables} lists dynamics quantities involved in the governing equations of the bone cells and bone porosity, Eqs~\eqref{obp}--\eqref{oca}, \eqref{fvas}. Table~\ref{table:biochem-parameters} lists all the parameters relevant to the biochemical regulation of the model. Table~\ref{table:biomech} lists quantities involved in the biomechanical regulation of the model. 

\paragraph{Concentrations of the biochemical signalling molecules}
As in Ref.~\cite{pivonka-etal-1,pivonka-etal-2,scheiner-etal-mechanostat,buenzli-pivonka-smith}, the concentration of the biochemical signalling molecules are governed by rate equations based on mass action kinetics. Ligand--receptor binding reactions occur on a time scale much faster than the characteristic times of cellular response (such as differentiation, apoptosis). The rate equations for the biochemical signalling molecules can therefore be taken in their steady state (see Refs.~\cite{pivonka-etal-1,buenzli-pivonka-smith} for details). This gives:
\begin{align}
    &\tgfb(t) = n_\tgfb^\text{bone} \kres \oca(t)/\tilde D_\tgfb \label{tgfb}
    \\&\rank(t) = N^\rank_\ocp\ \ocp(t), \label{rank}
    \\&\opg(t) = \frac{\beta^\opg_\oba\ \oba(t)\, \pi^\pth_{\text{rep},\ob}}{\beta^\opg_\oba\ \oba(t)\, \pi^\pth_{\text{rep},\ob}/\opg_\text{max} + \tilde D_\opg} \label{opg},
    \\&\rankl(t) = \frac{\rankl_\text{eff}\ \Bigg( \frac{\textstyle\beta_\rankl + P_\rankl^{\,\sed}}{\textstyle\beta_\rankl + \tilde{D}_\rankl \rankl_\text{eff}}\Bigg)} {1 + K_{[\rankl\text{--}\opg]}\opg + K_{[\rankl\text{--}\rank]} \rank}
    \label{rankl}.
    \\&\pth(t) = \big[P_\pth(t) + \beta_\pth\big]/\tilde D_\pth \label{pth}.
\end{align}
In Eq.~\eqref{rankl}, $\rankl_\text{eff}$ is the maximum concentration of \rankl\ (also referred to as effective carrying capacity), which is regulated by the parathyroid hormone $\pth$:
\begin{align}
    \rankl_\text{eff} = N^\rankl_\obp\ \obp\ \pi^\pth_{\text{act,\ob}},
\end{align}
where $N^\rankl_\obp$ is the maximum number of membrane-bound \rankl\ molecules that can be expressed on a single pre-osteoblast (see Refs.~\cite{pivonka-etal-1,pivonka-etal-2} for more details). In this work, the concentration of \mcsf, and so the quantity $\pi^\mcsf_{\text{act},\ocu}$, are assumed constant (see below and Table~\ref{table:biochem-parameters}). The additional production rate of \pth\ in Eq.~\eqref{pth}, $P_\pth(t)$, is used to increase the normal systemic levels of \pth\ to simulate an osteoporotic condition (see comments in Section~\ref{sec:results}).

\paragraph{Activator and repressor functions}
The regulation of the bone cell behaviours (such as differentiation, apoptosis, expression rate of a ligand) by the biochemical signalling molecules is modelled in Eqs~\eqref{obp}--\eqref{oca} by so-called ``activator'' functions $\pi^L_{\text{act},A}$ and ``repressor'' functions $\pi^L_{\text{rep}, A}$, where $L$ denotes the signalling molecule (ligand) and $A$ the signalled cell. These activator and repressor functions represent the strength of the response of the cell to the signal mediated by the ligand and are assumed to be given by:
\begin{align}
    \pi^L_{\text{act}, A} = \frac{L}{L+k^L_A }, \qquad\pi^L_{\text{rep},A} = 1-\pi^L_{\text{act},A} = \frac{1}{L + k^L_A},\label{piact-pirep}
\end{align}
where $k^L_A$ is the dissociation binding constant between the ligand and its receptor on the cell. The quantity $\pi^L_{\text{act,A}}$ represents the fraction of the receptors on the cell $A$ that are bound to a ligand $L$ (see Refs.~\cite{lemaire-etal,pivonka-etal-1} for more details). For example, the activator functions $\pi^\rankl_{\text{act},\ocu}$ and $\pi^\rankl_{\text{act},\ocp}$ are defined as:
\begin{align}
    \pi^\rankl_{\text{act},\ocu} = \pi^\rankl_{\text{act},\ocp} = \frac{\rankl}{\rankl + k^\rankl_\oc }.\label{pirankl}
\end{align}
where $k^\rankl_\oc$ is the dissociation binding constant between \rankl\ and the \rank\ receptor on \ocu s and \ocp s, and $\rankl$ is the free (unbound) $\rankl$ concentration given by Eq.~\eqref{rankl}.

\paragraph{Steady-state values}
The geometrical and biomechanical regulations of the bone cells is normalised by the steady-state values. These values depend in particular on the initial bone porosity, and so are calculated prior to evolving the system.

\begin{table}[h!]
    \centering
    \vspace{-3mm}
    \caption{Dynamic quantities in the governing equations, Eqs~\eqref{obp}--\eqref{oca}, \eqref{fvas}}\label{table:dynamic-variables}
    \small
    \begin{tabularx}{\columnwidth}{@{}l@{}c@{\hspace{2mm}}X@{}}
        \toprule
        Symbol & Unit & Description
        \\\thickmidrule
        $\ocp$ & \pM & density of pre-osteoclasts
        \\$\oca$ & \pM & density of active osteoclats
        \\$\obp$ & \pM & density of pre-osteoblasts
        \\$\oba$ & \pM & density of active osteoblasts
        \\\midrule
        $\tgfb$ & \pM & concentration of transforming growth factor $\betaup$
        \\$\rank$ & \pM & concentration of receptor-activator nuclear factor $\kappa$B
        \\$\rankl$ & \pM & concentration of receptor-activator nuclear factor $\kappa$B ligand
        \\$\opg$ & \pM & concentration of osteoprotegerin
        \\$\pth$ & \pM & concentration of parathyroid hormone
        \\\midrule
        \fvas & -- & volume fraction of vascular pores
        \\\sv & -- & specific surface
        \\\midrule
        $\sed$ & MPa & microscopic strain energy density of the bone matrix
        \\\bottomrule
    \end{tabularx}
\end{table}

\begin{table}[t]
    \centering
        \vspace{-3mm}
    \caption{Biochemical parameters}\label{table:biochem-parameters}
  \small
        \begin{tabularx}{\columnwidth}{@{}l@{\hspace{-2mm}}r@{\hspace{2mm}}X@{}}
        \toprule
        Symbol & Value & Description
        \\\thickmidrule
        $\ocu$ & \num{1e-3}\ \pM & density of uncommitted osteoclast progenitors
        \\$\obu$ & \num{1e-3}\ \pM & density of uncommitted osteoblast progenitors
        \\$\kres$ & 200 $\pM^{-1}\da^{-1}$ & daily volume of bone matrix resorbed per osteoclast
        \\$\kform$ & 40 $\pM^{-1}\da^{-1}$ & daily volume of bone matrix formed per osteoblast
        \\\midrule
        $D_\ocu$ & $4.2/\da$& $\ocu\to\ocp$ differentiation rate parameter
        \\$D_\ocp$ & $2.1/\da$& $\ocp\to\oca$ differentiation rate parameter
        \\$A_\oca$ & $5.65/\da$& \oca\ apoptosis rate parameter
        \\$D_\obu$ & $0.7/\da$& $\obu\to\obp$ differentiation rate parameter
        \\$D_\obp$ & $0.166/\da$& $\obp\to\oba$ differentiation rate parameter
        \\$P_\obp$ & $0.021/\da$& \obp\ proliferation rate parameter
        \\$A_\oba$ & $0.111/\da$& \oba\ apoptosis rate
        \\\midrule
        $\pi^\mcsf_{\text{act}, \ocu}$ & 0.5& value of the activator function of \mcsf\ for $\ocu\to\ocp$ differentiation
        \\$k^\rankl_\oc$ &5.68 \pM& dissociation binding constant for \rankl\ binding on \ocu\ and \ocp
        \\$k^\tgfb_\oca$ &\num{5.63e-4}\ \pM& dissociation binding constant for \tgfb\ binding on \oca
        \\$k^\tgfb_\obu$ &\num{5.63e-4}\ \pM& dissociation binding constant for \tgfb\ binding on \obu
        \\$k^\tgfb_\obp$ &\num{1.75e-4}\ \pM& dissociation binding constant for \tgfb\ binding on \obp
        \\$k^\pth_{\ob,\text{act}}$ &150\ \pM& dissociation binding constant for \pth\ binding on \ob\ (in $\pi^\pth_{\text{act},\ob}$)
        \\$k^\pth_{\ob,\text{rep}}$ &0.222\ \pM& dissociation binding constant for \pth\ binding on \ob\ (in $\pi^\pth_{\text{rep},\ob}$)
        \\$k_\text{[\rankl--\rank]}$ &0.034/\pM& association binding constant for \rankl\ and \rank
        \\$k_\text{[\rankl--\opg]}$ &0.001/\pM& association binding constant for \rankl\ and \opg
        \\$\beta^\rankl$ &$\num{1.68e2}~\pM/\da$& production rate of \rankl
        \\$\beta_\pth$ &$250\,\pM/\da$& production rate of systemic \pth
        \\$P_\pth$ &$500\,\pM/\da$& continous administration rate of \pth\ to model osteoporosis
        \\$\beta^\opg_\oba$ &$\num{1.62e8}/\da$& production rate of \opg\ per \oba
        \\$N^\rankl_\obp$ & \num{2.7e6}& maximum number of \rankl\ per \obp
        \\$N^\rank_\ocp$ & \num{1e4}& number of \rank\ receptors per \ocp
        \\$\opg_\text{max}$ & $\num{2e8}\,\pM$ & \opg\ density at which endogeneous production stops
        \\$n^\text{bone}_\tgfb$ & $0.01\,\pM$& density of \tgfb\ stored in the bone matrix
        \\$\tilde D_\tgfb$ &$2/\da$& degradation rate of \tgfb
        \\$\tilde D_\rankl$ &$10/\da$& degradation rate of \rankl
        \\$\tilde D_\opg$ &$0.35/\da$& degradation rate of \opg
        \\$\tilde D_\pth$ &$86/\da$& degradation rate of \pth
        \\\bottomrule
    \end{tabularx}
\end{table}

\begin{table}
    \centering
    \caption{Biomechanical quantities}\label{table:biomech}
  \small
        \begin{tabularx}{\columnwidth}{@{}l@{}r@{\hspace{2mm}}X@{}}
        \toprule
        Symbol & Value & Description
        \\\thickmidrule
        $\boldsymbol{\Sigma}$ & $\Sigma^\text{uni} \textbf{e}_3 \otimes\textbf{e}_3,$& macroscopic stress tensor
        \\ & $\Sigma^\text{uni} = -30~\text{MPa}$&
        \\$\varepsilon_\bm$ & & microscopic strain tensor of the bone matrix
        \\$\sed$ & & microscopic strain energy density of the bone matrix
        \\\midrule
        $\kappa$ & \num{1e5}~$\pM/\da$ & strength of biomechanical transduction of bone resorption
        \\$\lambda$ & 0.1 \text{or} 0.5 & strength of biomechnical transduction of bone formation
        \\\bottomrule
        \end{tabularx}
\end{table}
\clearpage

\end{appendices}


\begin{thebibliography}{99}
    \small
    \setlength{\itemsep}{-0.5ex}
    \bibitem{taylor-hazenberg-lee} Taylor D, Hazenberg JG, Lee TC (2007). Living with cracks: Damage and repair in human bone. \textit{Nat. Mat.} \textbf{6}:263--268
    \bibitem{parfitt2-in-recker} Parfitt AM (1983). The physiological and clinical significance of bone histomorphometric data. In Recker RR (Ed.), \textit{Bone histomorphometry: Techniques and interpretation}. CRC Press, Boca Raton, pp. 143--223.
    \bibitem{martin-burr-sharkey} Martin RB, Burr DB and Sharkey NA (1998). \textit{Skeletal Tissue Mechanics} (New York: Springer)
    \bibitem{parfitt-1994} Parfitt A M (1994) Osteonal and hemi-osteonal remodeling: the spatial and temporal framework for signal traffic in adult human bone. \textit{J. Cell. Biochem.} \textbf{55}:273--286
    \bibitem{martin-rankl} Martin TJ (2004) Paracrine regulation of osteoclast formation and activity: Milestones in discovery. \textit{J. Musculoskel. Neuron. Interact.} \textbf{4}:243--253
    \bibitem{roodman} Roodman GD (1999) Cell biology of the osteoclast. \textit{Exp. Hematology} \textbf{27}:1229--1241
    \bibitem{frost-mechanostat1} Frost HM (1987). Bone ``mass'' and the ``mechanostat'': A proposal. \textit{Anat. Rec.} \textbf{219}:1--9
    \bibitem{frost-mechanostat2} Frost HM (1990). Skeletal structural adaptations to mechanical usage (SATMU): 2. Redefining Wolff's law: The remodelling problem. \textit{Anat. Rec.} \textbf{226}:414--422
    \bibitem{frost-mechanostat3} Frost HM (2003). Bone's mechanostat: A 2003 Update. \textit{Anat. Rec.} \textbf{275A}:1081--1101
    \bibitem{smit-burger} Smit TH and Burger EH (2000). Is \bmu-coupling a strain-regulated phenomenon? A finite element analysis. \textit{J. Bone Miner. Res.} \textbf{15}:301--307
    \bibitem{burger-klein-nulend-smit} Burger EH, Klein-Nulend J and Smit TH (2003) Strain-derived canalicular fluid flow regulates osteoclast activity in a remodelling osteon---a proposal. \textit{J. Biomech.} \textbf{36}:1453--1459
    \bibitem{lee-staines-taylor} Lee TC, Staines A and Taylor D (2002) Bone adaptation to load: microdamage as a stimulus for bone remodelling. \textit{J. Anat.} \textbf{201}:437--446
    \bibitem{martin-1984} Martin RB (1984). Porosity and specific surface of bone. \textit{Crit. Rev. Biomed. Eng.} \textbf{10}:179--222         
    \bibitem{hellmich-etal} Hellmich C, Ulm F-J and Dormieux L (2004). Can the diverse elastic properties of trabecular and cortical bone be attributed to only a few tissue-independent phase properties and their interactions? \textit{Biomechan. Model. Mechanobiol.} \textbf{2}:219--238
    \bibitem{fritsch-hellmich} Fritsch A and Hellmich C (2007). ``Universal'' microstructural patterns in cortical and trabecular, extracellular and extravascular bone materials: Micromechanics-based prediction of anisotropic elasticity. \textit{J. Theor. Biol.} \textbf{244}:597--620
    \bibitem{grimal-etal} Grimal Q, Raum K, Gerisch A and Laugier P (2011). A determination of the minimum sizes of representative volume elements for the prediction of cortical bone elastic properties. \textit{Biomech. Model. Mechanobiol.} \textbf{10}:925--937
    \bibitem{martin-1972} Martin RB (1972). The effects of geometrical feedback in the development of osteoporosis. \textit{J. Biomech.} \textbf{5}:447--455
    \bibitem{nishiyama-etal} Nishiyama KK, Macdonald HM, Buie HR, Hanley DA and Boyd SK (2010). Postmenopausal women with osteopenia have higher cortical porosity and thinner cortices at the distal radius and tibia than women with normal aBMD: an in vivo HR-pQCT study. \textit{J. Bone Miner. Res.} \textbf{25}:882--890
    \bibitem{cooper-etal-2007} Cooper DML, Thomas CDL, Clement JG, Turinsky AL, Sensen CW and Hallgr\'imsson B (2007). Age-dependent change in the 3D structure of cortical porosity at the human femoral midshaft. \textit{Bone} \textbf{40}:957--965
    \bibitem{stevenson-marsh} Stevenson JC and Marsh M (2007). \textit{An atlas of osteoporosis}, 3rd Ed. (London, Informa Healthcare)
    \bibitem{hazelwood-etal} Hazelwood SJ, Martin RB, Rashid MM and Rodrigo JJ (2001). A mechanistic model for internal bone remodeling exhibits different dynamic responses in disuse and overload. \textit{J. Biomech.} \textbf{34}:299--308
    \bibitem{scheiner-etal-mechanostat} Scheiner S, Pivonka P, \etal (2012). Mechanobiological regulation of bone remodeling---Theoretical development of a coupled systems biology--micromechanical approach. Preprint: {\small \href{http://arxiv.org/abs/1201.2488}{arXiv:1201.2488}}
    \bibitem{pivonka-etal-1} Pivonka P, Zimak J, Smith DW, Gardiner BS, Dunstan CR, Sims NA, Martin TJ and Mundy GR (2008). Model structure and the control of bone remodeling: A theoretical study. \textit{Bone} \textbf{43}:249
    \bibitem{pivonka-etal-2} Pivonka P, Zimak J, Smith DW, Gardiner BS, Dunstan CR, Sims NA, Martin TJ and Mundy GR (2010). Theoretical investigation of the role of the \rank--\rankl--\opg\ system in bone remodeling. \textit{J. Theor. Biol.} \textbf{262}:306--316
    \bibitem{lemaire-etal} Lemaire V, Tobin FL, Greller LD, Cho CR, Suva LJ (2004). Modeling the interactions between osteoblast and osteoclast activities in bone remodeling. \textit{J. Theor. Biol.} \textbf{229}:293--309
    \bibitem{buenzli-pivonka-smith} Buenzli PR, Pivonka P, Smith DW (2011). Spatio-temporal structure of cell distribution in Bone Multicellular Units: A mathematical model. \textit{Bone} \textbf{48}:918--926
    \bibitem{wang-etal-bone-mm} Wang Y, Pivonka P, Buenzli PR, Smith DW and Dunstan CR (2011). Computational Modeling of Interactions between Multiple Myeloma and the Bone Microenvironment. \textit{PLoS ONE} \textbf{6}(11): e27494
    \bibitem{scheiner-etal-denosumab} Scheiner S, Pivonka P, and Dunstan CR (2012) Computer simulation-based modeling of the pharmaceutical intervention of postmenopausal osteoporosis by denosumab. To appear in the Proceedings of the 11th Biennal Conference on Engineering Systems Design and Analysis ASME 2012, July 2012, Nantes, France
    \bibitem{lauffenburger-linderman} Lauffenburger DA and Linderman JJ (1993). \textit{Receptors: models for binding, trafficking, and signaling.} (New York: Oxford Univ. Press)                
    \bibitem{bonewald-johnson} Bonewald LF and Jonhson ML (2008). Osteocytes, mechanosensing and \wnt\ signaling. \textit{Bone} \textbf{42}:606--615
    \bibitem{dullien} Dullien FAL (1992). \textit{Porous media. Fluid transport and pore structure},  2nd Ed. (San Diego: Academic Press)
    \bibitem{oers-etal} van Oers, RFM, Ruimerman R, Tanck E, Hilbers PAJ, and Huiskes R (2008). A unified theory for osteonal and hemi-osteonal
  remodeling. \textit{Bone} \textbf{42}:250--259
    \bibitem{hellmich-kober-erdmann} Hellmich C, Kober C and Erdmann B (2008). Micromechanics-based conversion of CT data into anisotropic elasticity tensors, applied to FE simulations of a mandible. \textit{Ann. Biomed. Eng.} \textbf{36}:108--122
    \bibitem{frolik-etal} Frolik CA, Black EC, Cain RL, Satterwhite JH, Brown-Augsburger PL, Sato M and Hock JM (2003). Anabolic and catabolic bone effects of human parathyroid hormone (1-34) are predicted by duration of hormone exposure.  \textit{Bone} \textbf{33}:372--379
    \bibitem{burr-etal} Burr DB, Hirano T, Turner CH, Hotchkiss C, Brommage R, and Hock JM (2001). Intermittently administered human parathyroid hormone(1-34) treatment increases intracortical bone turnover and porosity without reducing bone strength in the humerus of ovariectomized cynomolgus monkeys. \textit{J. Bone Miner. Res.} \textbf{16}:157-165
    \bibitem{ma-etal} Ma Y, Jee WS, Yuan Z, Wei W, Chen H, Pun S, Liang H and Lin C (1999). Parathyroid hormone and mechanical usage have a synergistic effect in rat tibial diaphyseal cortical bone.  \textit{J. Bone Miner. Res.} \textbf{14}:439-448
    \bibitem{turner-etal} Turner RT, Lotinun S, Hefferan TE, and Morey-Holton E (2006). Disuse in adult male rats attenuates the bone anabolic response to a therapeutic dose of parathyroid hormone.  \textit{J. Appl. Physiol.} \textbf{101}:881-886
    \bibitem{mayhew-clement-etal} Mayhew PM, Thomas CDL, Clement JG, Loveridge N, Beck TH, Bonfield W, Burgoyne CJ and Reeve J (2005). Relation between age, femoral neck cortical stability, and hip fracture risk. \textit{Lancet} \textbf{366}:129--135
    \bibitem{thomas-feik-clement} Thomas CDL, Feik SA and Clement JG (2005). Regional variation of intracortical porosity in the midshaft of the human femur: age and sex differences. \textit{J. Anat.} \textbf{206}:115--125
    \bibitem{seeman} Seeman E (2002). Osteoporosis II. Pathogenesis of bone fragility in women and men. \textit{Lancet} \textbf{359}:1841--1850
    \bibitem{parfitt1-in-recker} Parfitt AM (1983). Stereological basis of bone histomorphometry: Theory of quantitative microscopy and reconstruction of the third dimension. In Recker RR (Ed.), \textit{Bone histomorphometry: Techniques and interpretation}. CRC Press, Boca Raton, pp. 53--87.
\end{thebibliography}
\end{document}